\documentclass{emulateapj}

\slugcomment{}

\shorttitle{The JWST fluxes and colours of pop III galaxies}
\shortauthors{Zackrisson et al.}

\begin{document}

\title{The spectral evolution of the first galaxies. I.\\ JWST detection limits and colour criteria for population III galaxies}

\author{Erik Zackrisson\altaffilmark{1,2}$^*$, Claes-Erik Rydberg\altaffilmark{1,2}, Daniel Schaerer\altaffilmark{3,4}, G\"oran \"Ostlin\altaffilmark{1,2}, Manan Tuli\altaffilmark{1,2}}
\altaffiltext{*}{E-mail: ez@astro.su.se}
\altaffiltext{1}{Department of Astronomy, Stockholm University, 10691 Stockholm, Sweden}
\altaffiltext{2}{Oskar Klein Centre for Cosmoparticle Physics, AlbaNova University Centre, 10691 Stockholm, Sweden}
\altaffiltext{3}{Geneva Observatory, University of Geneva, 51 Chemin des Maillettes, CH-1290 Versoix, Switzerland}
\altaffiltext{4}{CNRS, IRAP, 14 Avenue E. Belin, F-31400 Toulouse, France}

\begin{abstract}
The James Webb Space Telescope (JWST) is expected to revolutionize our understanding of the high-redshift Universe, and may be able to test the prediction that the first, chemically pristine (population III) stars formed with very high characteristic masses. Since isolated population III stars are likely to be beyond the reach of JWST, small population III galaxies may offer the best prospects of directly probing the properties of metal-free stars. Here, we present Yggdrasil, a new spectral synthesis code geared towards the first galaxies. Using this model, we explore the JWST imaging detection limits for population III galaxies and investigate to what extent such objects may be identified based on their JWST colours. We predict that JWST should be able to detect population III galaxies with stellar population masses as low as $\sim 10^5\ M_\odot$ at $z\approx 10$ in ultra deep exposures. Over limited redshift intervals, it may also be possible to use colour criteria to select population III galaxy candidates for follow-up spectroscopy. The colours of young population III galaxies dominated by direct star light can be used to probe the stellar initial mass function (IMF), but this requires almost complete leakage of ionizing photons into the intergalactic medium. The colours of objects dominated by nebular emission show no corresponding IMF sensitivity. We also note that a clean selection of population III galaxies at $z\approx 7-8$ can be achieved by adding two JWST/MIRI filters to the JWST/NIRCam filter sets usually discussed in the context of JWST ultra deep fields. 
\end{abstract}





\section{Introduction}
\label{intro}
Due to the lack of efficient coolants in metal-free, primordial gas, the first stars to form in our Universe were likely very massive \citep[$\gtrsim 10\ M_\odot$ and possibly even up to 1000$\ M_\odot$; e.g.][]{Bromm et al. a,Nakamura & Umemura,Tan & McKee,Greif & Bromm,Ohkubo et al.}. Some of these population III stars probably exploded as supernovae \citep{Heger et al.} and initiated the chemical enrichment which allowed less massive and more metal-rich stars to form (the population I and II stars known from the local Universe, with characteristic masses below $1\ M_\odot$). The highly energetic radiation emitted from population III stars may have played an important role in the cosmic reionization at redshifts $z>6$ \citep[e.g.][]{Sokasian et al.,Trenti & Stiavelli b}, and the remnants left behind by these objects may also have acted as seeds for the supermassive black holes \citep[e.g.][]{Trenti & Stiavelli a} known to exist already at $z\approx 6$ \citep[e.g.][]{Willott et al.}. 

Current predictions for the initial mass function of population III stars are, however, still shaky. The most recent simulations suggest that the typical population III masses may be lower than previously believed \citep[e.g.][]{Stacy et al.,Clark et al.,Greif et al. d,Greif et al. e}, and the chemical abundance patterns of extremely metal-poor stars in the Milky Way halo also point towards early supernovae with progenitor masses closer to $\sim 10\ M_\odot$ than $\sim 100\ M_\odot$ \citep[for reviews, see][]{Beers & Christlieb,Karlsson et al. b}. At the same time, it is not clear that the chemical signatures of the most massive population III supernovae would be detectable in current halo samples. The stars bearing the mark of such supernovae may be too scarce \citep[e.g.][]{Salvadori et al.,Trenti & Shull}, hiding at too high metallicities \citep{Karlsson et al. a} or in the inner regions of the Galaxy \citep[e.g.][]{Tumlinson}. Direct observations of massive population III stars in the high-redshift Universe would help settle the issue, but this is beyond the capabilities of current telescopes.

The very first population III stars are expected to form in isolation or in small numbers within $\sim$10$^5$--$10^6 M_\odot$ dark matter halos at redshifts $z\approx 20$--50 \citep[e.g.][]{Tegmark et al.,Yoshida et al.}, but the prospects of detecting such stars on an individual basis appear bleak \citep[e.g.][]{Gardner et al.,Greif et al. b,Rydberg et al.}, at least before they go supernovae \citep{Weinmann & Lilly,Whalen & Fryer}. However, population III stars may continue to form within the more massive halos ($\gtrsim 10^{7-8} \ M_\odot$) hosting the first galaxies\footnote{Due to their low expected stellar population masses, these objects are sometimes also -- and perhaps more appropriately -- referred to as population III star clusters} at $z\lesssim 15$ \citep{Scannapieco et al.,Schneider et al. a,Tornatore et al.,Johnson et al. a,Johnson et al. b,Stiavelli & Trenti,Johnson}, and this could in principle allow their integrated signatures to be detected with the upcoming {\it James Webb Space Telescope}\footnote{http://www.jwst.nasa.gov/} \citep{Bromm et al. b,Johnson et al. b,Johnson}. 

Since the first generation of galaxies is predicted to form in high-density regions that have been pre-enriched by pop III stars in minihalos, these galaxies are not expected to be metal-free, and are most likely dominated by chemically enriched stars \citep[e.g.][]{Greif et al. c}. True population III galaxies may, however, form at slightly later epochs, in low-density environments which have remained chemically pristine \citep[e.g.][]{Scannapieco et al.,Trenti et al.,Stiavelli & Trenti}. 

Due to the exceedingly high temperatures of massive population III stars ($T_\mathrm{eff}\sim 10^5$ K), such objects may contribute substantially to the photoionization of the interstellar medium and alter the overall spectra of their host galaxies. A number of authors have pointed out that population III-dominated galaxies could be identified based on the strengths of the Ly$\alpha$ line, the HeII $\lambda1640$, $\lambda4686$ lines, or the Lyman `bump' \citep[e.g.][]{Tumlinson & Shull,Tumlinson et al. a,Oh et al.,Malhotra & Rhoads,Schaerer a,Schaerer b,Inoue} and several searches have since tried to apply these techniques \citep[e.g.][]{Dawson et al.,Nagao et al. a,Dijkstra & Wyithe,Ouchi et al.,Nagao et al. c,Cai et al.}. While a few potential population III candidates have indeed been reported \citep{Fosbury et al.,Raiter et al. a,di Serego Alighieri,Inoue et al.}, their exact nature remain unclear. So far, essentially all searches for population III galaxies have relied on spectroscopy or narrowband photometry. If population III galaxy candidates could instead be singled out from large samples based just on their broadband colours \citep[as recently attempted by][]{Bouwens et al.}, this would allow for substantial gains in terms of observing time. 

Here, we present {\it Yggdrasil}, a new spectral synthesis model for the first generations of galaxies, and use it to explore the spectral signatures of pure population III galaxies in {\it James Webb Space Telescope} (JWST) photometric surveys. The technical details of the model are described in Sect.~\ref{Yggdrasil}. In Sect.~\ref{scenarios}, we define three different classes of first galaxies, which differ by the relative amount that nebular emission is expected to contribute to their integrated spectra. The resulting {\it James Webb Space Telescope} (JWST) mass detection limits for both population III and more chemically evolved galaxies in photometric surveys are presented in Sect.~\ref{masslimits}. Potential strategies for searching for population III galaxies (dominated either by nebular emission or direct star light) in deep JWST images are presented in Sect.~\ref{typeA} and Sect.~\ref{typeC}. A number of potential caveats with the proposed approaches are discussed in Sect.~\ref{discussion}. Sect.~\ref{summary} summarizes our findings.
\begin{figure*}
\plottwo{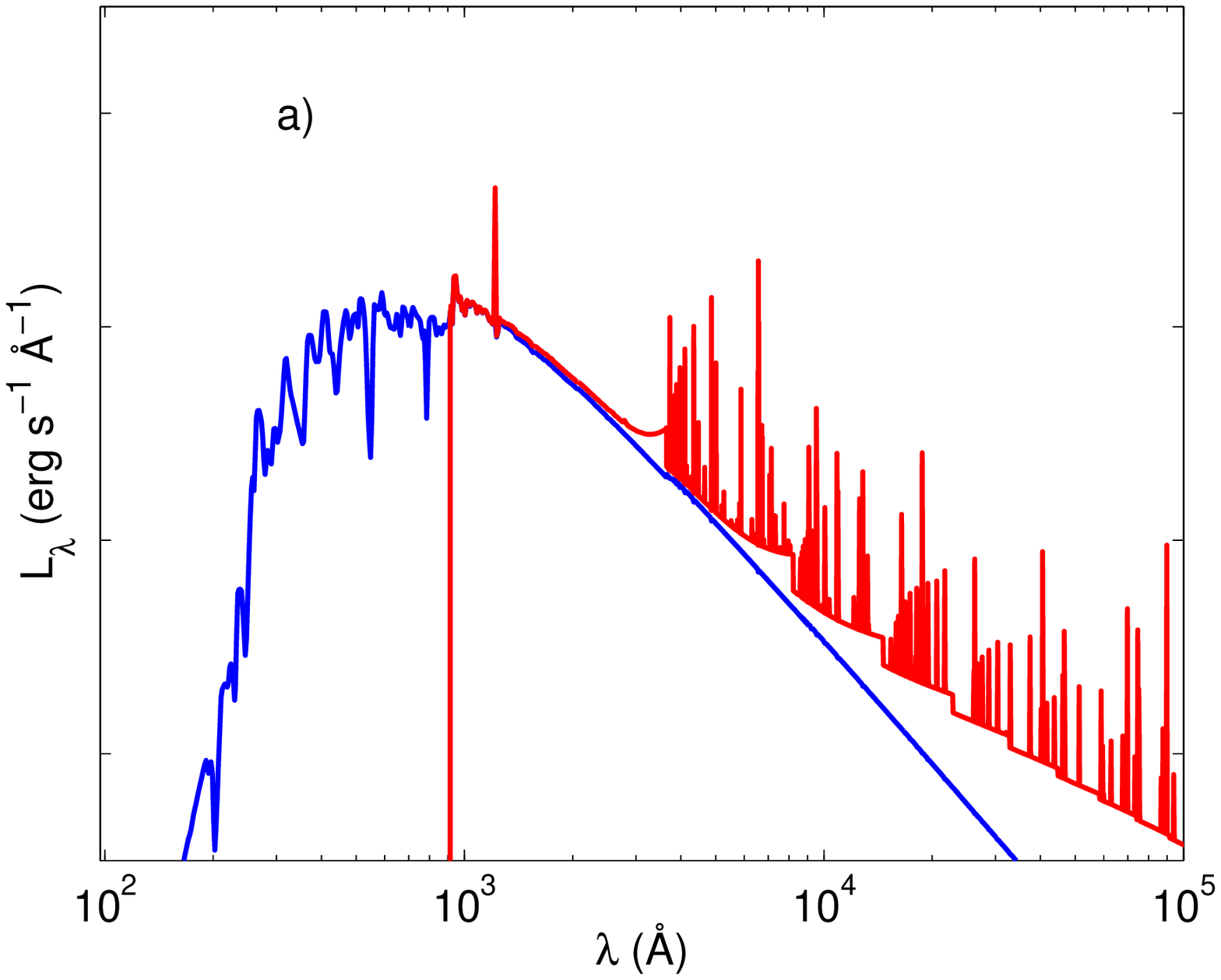}{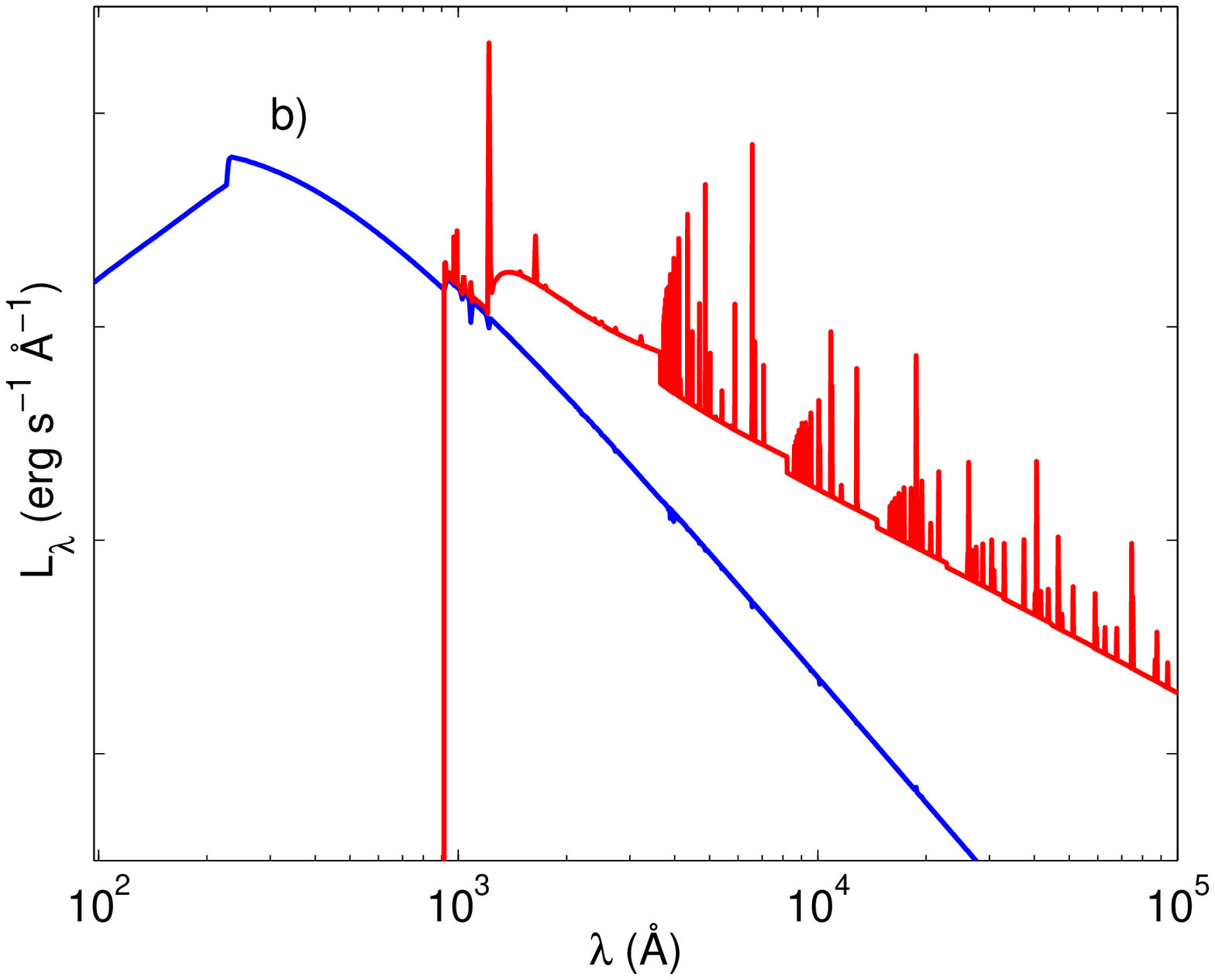}
\caption{The nebular contribution to the rest-frame SEDs of young, dust-free {\bf a)} pop I ($Z=0.020$) and {\bf b)} pop III.1 galaxies. Blue lines represent the purely stellar SEDs, whereas the red lines represent the total SEDs (including both stellar and nebular contributions).  An instantaneous burst and an age of 1 Myr is assumed in both cases.  The nebular component completely transforms the appearance of the SED in both panels, but the relative contribution of nebular emission becomes much higher for the pop III.1 object (right panel) because of a higher fraction of hot, high-mass stars, which boost the Lyman continuum at $\lambda < 912$\AA. The stellar SEDs have been rescaled to the same (arbitrary) flux level at 1300 \AA{} to facilitate a comparison between the panels. 
\label{spectra}}
\end{figure*}

\section{The Yggdrasil model}
\label{Yggdrasil}
While the properties of galaxies at $z\gtrsim 10$ remain an uncharted territory observationally, theory and simulations provide a number of clues as to what one might expect from such objects. In the models of \citet{Stiavelli & Trenti}, the very first galaxies are predicted to form in high-density regions that have been pre-enriched by population III stars that exploded as supernovae in minihalos at even higher redshifts. True population III galaxies start to form somewhat later, in low-density environments which have remained chemically pristine. Since pockets of primordial gas may survive in galaxies that have already experienced some chemical enrichment, hybrid galaxies in which population III, population II (representative of the metallicity in the Milky Way halo; $Z<Z_\odot/10$) and population I (representative of the metallicity in the Milky Way disk; $Z\gtrsim Z_\odot/10$) stars continue to form in parallel can also be expected \citep[e.g.][]{Salvaterra et al.}, and signatures of this may already have been detected \citep{Jimenez & Haiman}. Even more exotic galaxies may be envisioned if the dark matter of the Universe has the properties required for the formation of long-lived population III stars fueled by WIMP annihilations in minihalos \citep[e.g.][]{Spolyar et al.}. Many such ``dark stars'' may then end up within the first galaxies during their hierarchical assembly, and possibly imprint detectable signatures in their spectra \citep{Zackrisson et al. c}. In this paper, however, we focus on pure population III galaxies.

Here, we present Yggdrasil\footnote{named after a sacred tree in Norse mythology}, a new population synthesis code custom-designed for modelling the spectral energy distributions (SEDs) of high-redshift galaxies. To reflect the significant variance in terms of stellar content that these objects may display, Yggdrasil is equipped to handle mixtures of conventional population I, II and III stars (hereafter pop I, II and III) as well as dark stars. This code also allows the treatment of nebular emission from the photoionized gas and extinction due to dust. A number of Yggdrasil model grids are publicly available from the lead author's homepage\footnote{www.astro.su.se/$\sim$ez}. 

The SEDs of Single Stellar Populations (SSPs\footnote{also known as single-age stellar populations or instantaneous-burst populations}), from various other population synthesis models can be used as input to Yggdrasil, which will then reweight the SSP time steps to accommodate arbitrary star formation histories. Currently, the SSP models of \citet{Zackrisson et al. a}, Starburst99 \citep{Leitherer et al.,Vazquez & Leitherer} and \citet{Maraston} are implemented as options for pop I and II stars. For the duration of this paper, we will adopt Starburst99 SSPs generated with Padova-AGB stellar evolutionary tracks \citep{Vazquez & Leitherer} and the \citet{Kroupa} universal stellar initial mass function (IMF) throughout the mass range 0.1--100 $M_\odot$ for population II (assumed metallicity $Z=0.0004$) and pop I (assumed metallicity $Z=0.020$) galaxies. 

While both theoretical arguments and numerical simulations give strong support to the notion that pop III stars must have been more massive than the pop II and I stars forming later on \citep[for a review, see][]{Bromm & Larson}, the exact IMF of pop III stars remains unknown. It has been argued that the Universe may have produced two classes of pop III stars -- pop III.1 stars which formed first, with characteristic masses of about $\sim 100\ M_\odot$, and pop III.2 stars which formed somewhat later and had lower characteristic masses of $\sim 10\ M_\odot$ due to HD cooling promoted by the Lyman-Werner feedback provided by the pop III.1 stars \citep[e.g.][]{Mackey et al.,Greif & Bromm}. Naively, one would expect the pop III.1 IMF to be appropriate for the stars forming in isolation (or in small numbers) within $\sim 10^{5-6}\ M_\odot$ minihalos capable of H$_2$ cooling at $z\approx 20$--40, whereas the pop III.2 IMF may be more relevant for the first pop III galaxies forming in $\gtrsim 10^{7-8}\ M_\odot$ halos capable of HI cooling at $z\leq 15$ \citep[e.g.][]{Johnson et al. a}. However, this pop III.1 ($\sim 100\ M_\odot$) versus pop III.2 ($\sim 10\ M_\odot$) picture is looking increasingly uncertain, as the latest simulations indicate more fragmentation than before, with lower pop III masses as a result \citep[e.g.][]{Stacy et al.,Clark et al.,Greif et al. d}. It has recently also been recognized that supersonic streaming velocities of baryons with respect to the dark matter \citep[e.g][]{Tseliakhovich & Hirata,Maio et al.} could have increased the turbulence within minihalos and assisted in reducing the masses of pop III.1 stars \citep{Greif et al. e}. For simplicity, we have chosen to stick to the traditional pop III.1 ($\sim 100\ M_\odot$) and III.2 ($\sim 10\ M_\odot$) convention throughout this paper. To cover all the bases, we adopt both of these top-heavy IMFs as viable options for our primordial galaxies, but also consider pop III galaxies with the same \citet{Kroupa} IMF as that used for pop II/I galaxies.

For pop III.1 galaxies, we will adopt the \citet{Schaerer a} stellar SSP with a power-law IMF ($\mathrm{d}N/d\mathrm{M}\propto M^{-\alpha}$) of slope $\alpha=2.35$ throughout the mass range 50--500 $M_\odot$. For pop III.2 galaxies, we will adopt the \citet{Raiter et al. b} TA model, which has a log-normal IMF with characteristic mass $M_\mathrm{c}=10\ M_\odot$, dispersion $\sigma=1\ M_\odot$ but wings extending from 1--500 $M_\odot$. 

The contribution to the SED from photoionized gas is computed using the procedure outlined in \citet{Zackrisson et al. a}. In this machinery, the stellar population SED is, at every age step (once the star formation history has been taken into account), used as input to the photoionization code Cloudy \citep{Ferland et al.}. This results in a self-consistent prediction for the nebular continuum and emission line fluxes which reflects the temporal changes in the ionizing radiation field. Other approaches in the literature include calculating the nebular continuum based on tabulated emissivities assuming a fixed electron temperature \citep[e.g.][ see \citealt{Raiter et al. b} for a description of the shortcomings of this technique]{Leitherer et al.}, or using emission-line ratios fixed by empirical calibrations \citep[e.g.][]{Anders & Fritze-v. Alvensleben}. While the latter technique may be useful for certain classes of objects, it will of be of little use for objects with properties that differ significantly from the empirical templates, since the emission-line ratios are expected to vary as a function of age, IMF, metallicity and the physical conditions in the nebula. Throughout this paper, we will assume the nebula to be spherical, ionization-bounded and to have a constant, non-evolving hydrogen density $n(\mathrm{H})$. These assumptions concerning the properties of the nebular gas, and their impact on our conclusions, are further discussed in Sect.~\ref{scenarios}. Additional Cloudy parameters include the gas filling factor $f_\mathrm{fill}$ (which describes the porosity of the gas), the gas covering factor $f_\mathrm{cov}$ (which regulates the amount of Lyman continuum leakage into intergalactic space) and the gas metallicity $Z_\mathrm{gas}$.

In the case of plane-parallel nebulae, there is a homology relation between photoionization models with the same ionization parameter $U$:
\begin{equation}
U=\frac{Q(\mathrm{H})}{4\pi c R^2 n(\mathrm{H})},
\end{equation}
where $Q(\mathrm{H})$ is the number of ionizing photons per unit time and $R$ is the distance to the ionization source. The situation becomes more complicated for spherical nebulae (as assumed here), due to the dilution of the radiation as it propagates through the nebula (giving high $U$ close to the centre and low $U$ further out). Instead of specifying $U$ directly, the inner radius of the cloud, $R_\mathrm{in}$ and the mass available for star formation, $M_\mathrm{tot}$ are instead used as input to Yggdrasil. The latter is defined as the gas mass converted into stars during a star formation episode of duration $\tau$:
\begin{equation}
M_\mathrm{tot}=\int_0^\tau \mathrm{SFR}(t)\; \mathrm{d}t.
\label{Mtot_eq}
\end{equation}
While the shape of the stellar SED is set by the stellar SSP, the age and the star formation history, $M_\mathrm{tot}$ determines the overall scaling, and hence luminosity and $Q(\mathrm{H})$. Please note that $M_\mathrm{tot}$ does not represent the mass of the resulting nebula, which may either be higher or lower than $M_\mathrm{tot}$, depending on $Q(\mathrm{H})$ and the gas parameters ($n(\mathrm{H})$, $f_\mathrm{fill}$, $f_\mathrm{cov}$ and to smaller extent $Z_\mathrm{gas}$). 

As in the \citet{Zackrisson et al. a} model, the inner radius of the ionizing cloud is set to: 
\begin{equation}
R_\mathrm{in} = 100\ R_\odot \left(\frac{L}{L_\odot}\right)^{1/2},
\end{equation}
where $L$ represents the bolometric luminosity of the model galaxy. This gives an effective mean ionization parameter in the same range as observed in local HII regions \citep{Kewley & Dopita}, if $f_\mathrm{fill}=0.01$ and $n(\mathrm{H})=100$ cm$^{-3}$. Throughout this paper, we adopt $M_\mathrm{tot}=10^6\ M_\odot$, to reflect the likely stellar population masses of the faintest galaxies detectable with JWST (see Sect.~\ref{masslimits}). However, we have verified that the resulting JWST colours remain practically unchanged for objects that are a up to a factor of $\sim 10^3$ more massive.  

While dust extinction is not expected to be present during the very first star formation episode of pop III galaxies, it may well become important after a few Myr, when pair-instability supernovae from 140--$260 \ M_\odot$ stars or type II supernovae with $M<50 \ M_\odot$ \citep[mass limits valid in the absence of stellar rotation;][]{Heger et al.} release their nucleosynthesis yields into the surroundings. In pop II/I galaxies, extinction is likely to be relevant at all ages. To allow for the treatment of dust extinction, Yggdrasil allows a choice of four different dust-correction recipes: the Milky Way, LMC, SMC  \citep{Pei} and \citet{Calzetti} attenuation models. In the latter case, corrections are applied separately to the nebular and stellar contributions to the SED, so that the nebular emission experiences a higher extinction. All dust extinction corrections are applied after the nebular SED has been generated. This is equivalent to assuming that the dust is located outside the HII region, and hence does not affect the Lyman continuum flux prior to gas absorption. This is reasonable assumption, at least for young pop III galaxies experiencing a brief star formation episode, since no current models predict dust formation in the immediate vicinities of pop III stars.

\section{The nebular properties of high-redshift galaxies}
\label{scenarios}
\begin{figure*}
\plottwo{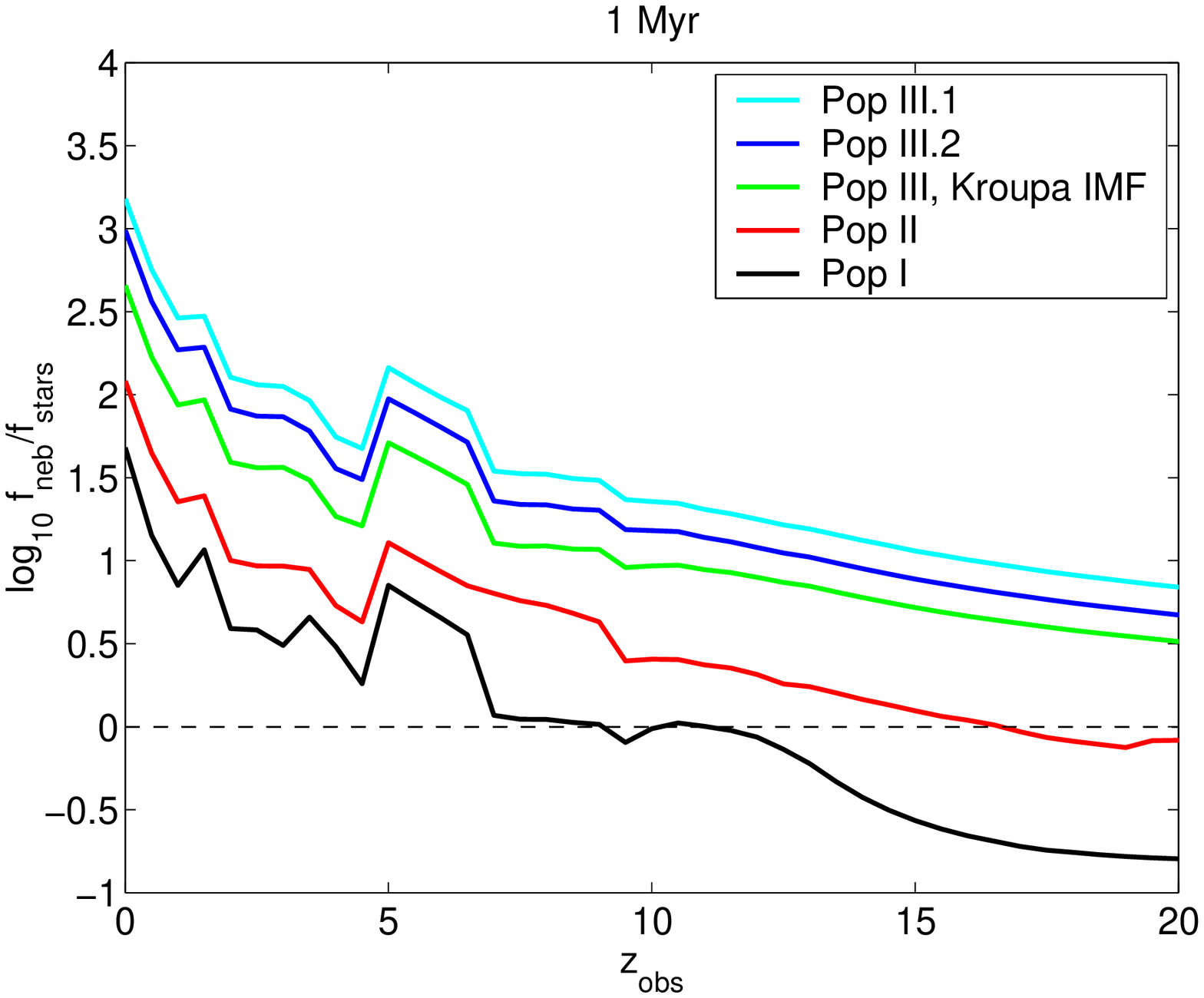}{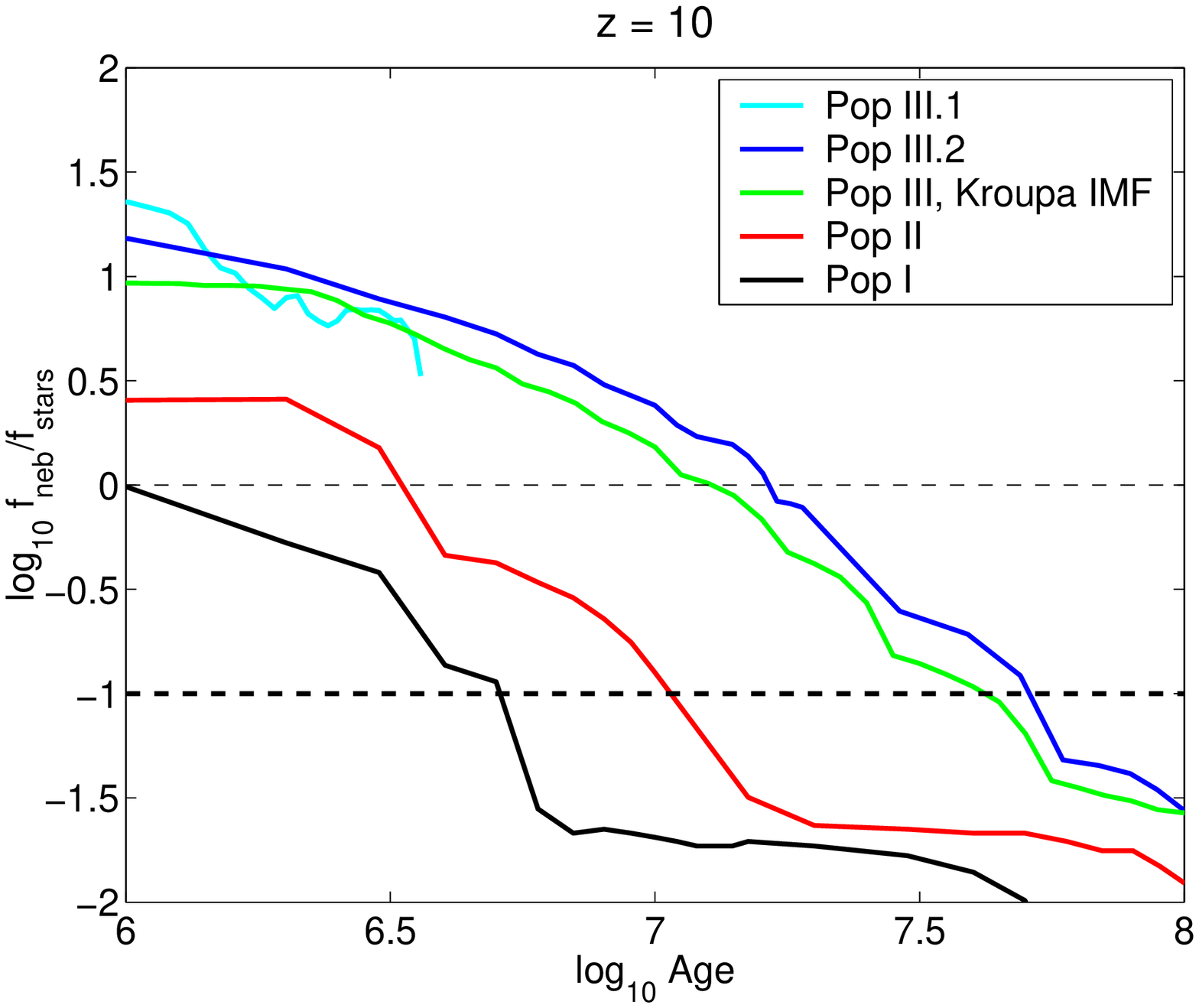}
\caption{The impact of nebular emission on pop I, II and III galaxies. {\bf a)} The ratio of nebular to stellar light as a function of redshift for a 1 Myr galaxy in the JWST/NIRCam F444W filter at 4.44 $\mu$m. The different colours correspond to different assumptions about the metallicity and IMF: Pop I (black), pop II (red), pop III with a \citet{Kroupa} IMF (green), pop III.2 (blue) and pop III.1 (cyan). The pop I and pop III.1 models correspond to the SEDs depicted in Fig.~\ref{spectra}. The bumps and wiggles along the lines are produced when various emission lines and continuum redshift in and out of the filter. Nebular emission is seen to dominate the observed fluxes ($\log_{10} f_\mathrm{neb}/f_\mathrm{stars} >0$; thin horizontal dashed line) at essentially all redshifts for Pop III and Pop II, but the exact contribution depends on the metallicity and IMF. The relative flux contribution from the nebula is always more than an order of magnitude higher in the population III.1 galaxy than in the population I galaxy. {\bf b)} The evolution of $f_\mathrm{neb}/f_\mathrm{stars}$ with age in the JWST/NIRCam F444W filter for instantaneous burst models. The line colours have the same meaning as in the left panel. The pop III.1 predictions (cyan line) ends abruptly at $\approx 3$ Myr, since no stars in the 50--$500 M_\odot$ range have life spans longer than this. In the case of the pop III.2 and pop III, \citet{Kroupa} IMF models, it takes $\approx 12$--16 Myr before direct stars start to dominate over nebular emission ($\log_{10} f_\mathrm{neb}/f_\mathrm{stars}<0$; thin horizontal dashed line), and $\approx 40$--50 Myr until nebular emission gives a negligible contribution ($\log_{10} f_\mathrm{neb}/f_\mathrm{stars}<-1$; thick horizontal dashed line) to the flux. In the case of young ($\leq 10$ Myr) pop III galaxies, it would take a very large amount of Lyman continuum leakage ($f_\mathrm{esc}\gtrsim 0.95$) to bring nebular emission down to a level where it no longer affects the broadband fluxes in any significant way. 
\label{fneb}}
\end{figure*}

Nebular emission typically has a strong impact on the SEDs of young or star-forming galaxies, and the relative contribution from nebular gas is expected to become stronger if the metallicity becomes very low or if the IMF turns more top-heavy \citep{Schaerer a,Schaerer b,Raiter et al. b}. This is demonstrated in Fig.~\ref{spectra}, where we compare the rest-frame SEDs of 1 Myr old SSP models for a pop I galaxy ($Z=0.020$, \citealt{Kroupa} IMF) and a pop III.1 galaxy. In both cases, the blue line represents the purely stellar SEDs whereas the red line describes the total SED, i.e. the SED including both stellar and nebular contributions. The nebular component completely transforms the appearance of the SED for both types of galaxies, but the relative contribution of nebular emission becomes much stronger for the zero-metallicity, top-heavy IMF object (pop III.1).

As we have argued in several previous papers \citep[e.g.][]{Zackrisson et al. b,Schaerer & de Barros a,Schaerer & de Barros b}, the flux boost due to photoionized gas can also have a notable impact on the broadband fluxes of high-redshift galaxies. In Fig.~\ref{fneb}a, we demonstrate this by plotting the ratio of nebular to stellar flux, $f_\mathrm{neb}/f_\mathrm{stars}$, in the JWST/NIRCam F444W filter (at 4.44 $\mu$m) as function of redshift (within the range $z=0$--20) for newborn (age 1 Myr), galaxies with SEDs representative of pop I ($Z=0.020$), pop II ($Z=0.0004$), pop III with \citet{Kroupa} IMF, pop III.2 and pop III.1 SEDs, assuming an instantaneous burst. Nebular emission is seen to dominate the flux in this filter at all redshifts considered for the pop III models. As expected, the highest $f_\mathrm{neb}/f_\mathrm{stars}$ are produced by the pop III.1 IMF (cyan line), followed by the pop III.2 IMF (blue line) and pop III with a \citet{Kroupa} IMF (green line). The pop II (red line) and pop I models (black line) produce significantly lower ionizing fluxes per unit mass.

The bumps and wiggles along the lines in Fig.~\ref{fneb}a are produced when various emission lines and continuum features redshift in and out of the filter. The bump at $z\approx 5$--7 is due to the H$\alpha$ line at 6563 \AA, and the extension of this bump to $z\approx 9$ in the case of the pop II galaxy is primarily due to the [OIII] line at 5007 \AA, which is strong at low to intermediate metallicities, weak in high-metallicity objects and completely missing from pop III spectra due to lack of metals in these objects, as further discussed in Sect.~\ref{typeA}. The overall increase in $f_\mathrm{neb}/f_\mathrm{stars}$ towards very low redshifts is due to the very different spectral slopes of the stellar and nebular continuum at rest-frame wavelengths above $10^4$ \AA{} (see Fig.~\ref{spectra}).

In Fig.~\ref{fneb}b, we plot the $f_\mathrm{neb}/f_\mathrm{stars}$ ratio at $z=10$ as a function of age for the same instantaneous-burst models as in Fig.~\ref{fneb}a. At an age of 1 Myr, going from a \citet{Kroupa} IMF pop I (black line) to a pop II (red line) increases the $f_\mathrm{neb}/f_\mathrm{stars}$ ratio by a factor of $\approx 2.5$, and going from a pop II to a pop III galaxy (green line) boosts $f_\mathrm{neb}/f_\mathrm{stars}$ by an additional factor of $\approx 4$. Shifting to more top-heavy IMFs (blue for pop III.2 and cyan for pop III.1) raises $f_\mathrm{neb}/f_\mathrm{stars}$ even further. These overall ratios between the models are approximately stable as a function of age, except for the pop III.1 model, which evolves much faster than the others due to the lack of stars with masses below $50\ M_\odot$, fading from sight after just 3 Myr. In the case of the pop III.2 and pop III, \citet{Kroupa} IMF models, it takes $\approx 12$--16 Myr before direct star light starts to dominate over nebular emission (thin horizontal dashed line), and $\approx 40$--50 Myr until nebular emission gives a negligible contribution (10\%, as indicated by the thick horizontal dashed line) to the flux.

Even though Lyman continuum leakage from high-redshift galaxies may well reduce the relative contribution from nebular emission (as further discussed below), it is clear from Fig.~\ref{fneb} that it would take a very large amount of leakage to break the dominance of nebular light in the SEDs of young pop III galaxies. For a 1 Myr pop III.1 galaxy, $f_\mathrm{neb}/f_\mathrm{stars}\approx 20$ at $z=10$, whereas the corresponding values for pop III.2 and pop III galaxies with a \citet{Kroupa} IMF are $f_\mathrm{neb}/f_\mathrm{stars}\approx 15$ and $\approx 10$ respectively. After 10 Myr, pop III galaxies with a pop III.2 or a \citet{Kroupa} IMF have $f_\mathrm{neb}/f_\mathrm{stars}\approx 5$ and $\approx 3$, respectively. To bring nebular emission down to a level where it no longer affects the broadband fluxes of $\leq 10$ Myr old pop III objects in any significant way ($f_\mathrm{neb}/f_\mathrm{stars}\sim 0.1$), the ionizing escape fraction would have to be $f_\mathrm{esc}\gtrsim 0.95$. 

It is doubtful whether pop III galaxies can be identified based on colour criteria for much longer than $\sim 10^7$ yr. The lifetimes of massive pop III stars are very short \citep[$\approx 2$--$3\times 10^6$ yr years for $\approx 50$--$500\ M_\odot$ stars; ][]{Schaerer a}, and as soon as the first supernovae explode, metals will be dispersed into the surrounding medium. While the ejecta from the first pop III supernovae may take $\sim 10^8$ yr to cool sufficiently to be used in the subsequent formation of pop II and pop I stars \citep{Greif et al. a}, the time it takes before metal emission lines start to emerge from the surrounding nebula may be much shorter, perhaps no more than a few Myr. At that point, the broadband signatures of metal-free nebulae discussed in Sect.~\ref{typeA} would be jeopardized. If feedback from the first supernovae clears the galaxy of photoionized gas, the SED would be dominated by direct star light from pop III stars until pop II/I star formation sets in (after $\sim 10^8$ yr), but even then, the unique spectral characteristics of pop III stars are only retained for $\sim 10^7$ yr (see Sect.~\ref{typeC}).

\subsection{Type A, B and C}
\begin{figure*}
\plotone{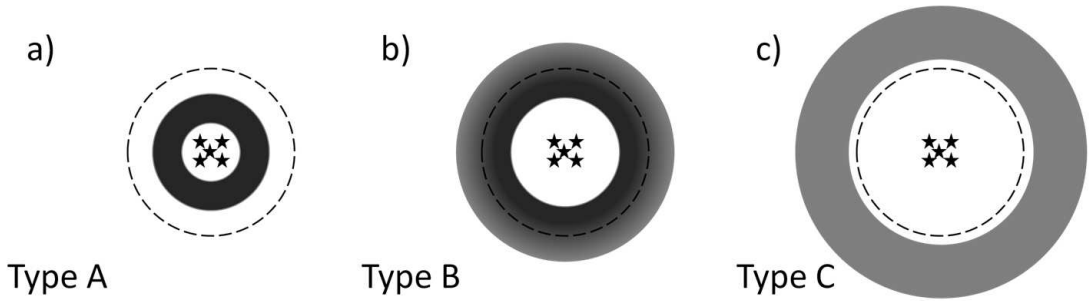}
\caption{Schematic illustration of the dynamical evolution of the ionized gas surrounding the first galaxies. {\bf a)} For type A galaxies, the HII region (dark region) is compact, ionization/radiation bounded and confined well within the virial radius (dashed line) of its dark matter halo. This implies a very low escape fraction for ionizing radiation ($f_\mathrm{esc}\approx 0$) and a maximal contribution of nebular emission to the overall SED of the galaxy. 
{\bf b)} In the case of type B, part of the HII region has been pushed outside the virial radius of the halo, with partial escape of ionizing radiation as a result ($0 < f_\mathrm{esc} < 1$). At this point, the SED may still contain signatures of nebular emission, but at a lower level than that provided for type A. {\bf c)} In type C objects, the HII region has been completely pushed outside the virial radius, implying a massive escape of ionizing photons into the intergalactic medium ($f_\mathrm{esc}\approx 1$). At this point, the nebula will be huge, have very low density, long recombination timescales and probably very little impact on the observations of individual galaxies. Hence, the SED is expected to be almost completely stellar, with essentially no contribution from nebular emission.
\label{schematic}}
\end{figure*}
While Fig.~\ref{spectra} and Fig.~\ref{fneb} would suggest that the SEDs of pop III galaxies are almost completely nebular in nature, there are a number of mechanisms that could diminish the impact of nebular emission. As shown in the simulations by \citet{Johnson et al. b}, stellar feedback can drive the photoionized gas present in these systems outward, eventually pushing it beyond the virial radius of the dark matter halo. Once the HII region breaks out of its host halo, the gas density will be extremely low, resulting in a huge intergalactic nebula with very long recombination timescales. This nebula, while very important for cosmic reionization, is probably of little importance for JWST observations of individual galaxies. When attempting aperture photometry of galaxies surrounded by such large, low-surface brightness nebulae, the nebular contribution (appearing in both the object and sky annulus) may be almost completely subtracted away, effectively leading to the detection of a SED dominated by the more concentrated stellar component. Whether or how quickly the gas gets expelled depends on the star formation efficiency within a given halo, and also the form of the IMF. Having a very low ratio of stellar to dark matter mass, and/or an IMF without too many high-mass stars, would likely result in a situation in which the HII region stays completely confined within the CDM halo \citep[for the limiting case of just one pop III star in halos with different masses, see e.g.][]{Kitayama et al.}.

In Fig.~\ref{schematic}, we have schematically defined three classes of objects in different stages of gas expulsion. For type A, the ionization-bounded HII region (grey region) remains compact and confined well inside the virial radius (dashed line). This implies a Lyman-continuum escape fraction $f_\mathrm{esc}\approx 0$ and a maximal contribution from nebular emission to the overall SED of the galaxy. In the case of type B, stellar feedback has pushed the HII-region partly outside the virial radius ($\sim 1$ kpc for a $10^8\ M_\odot$ halo at $z\approx 7$--13), giving rise to partial leakage of ionizing radiation into the intergalactic medium: $0 < f_\mathrm{esc} < 1$. For type C galaxies, the ionizing gas has been expelled outside the dark matter halo, resulting in an SED with minimal nebular contribution and a Lyman-continuum escape fraction $f_\mathrm{esc}\approx 1$. While this simple picture admittedly neglects the anisotropic outflow of gas and the irregularities within the photoionized medium evident from actual simulation \citep{Johnson et al. b}, it still captures the salient points when it comes to modelling the nebular contribution to the galaxy SED. 

In this paper, we will only treat galaxies that belong to categories A (full nebular contribution) and C (no nebular contribution). The intermediate type B would be far more challenging to model in detail -- the gas density profile is highly time-dependent at this stage, the long recombination timescales introduce a time lag between the evolution of the stellar and nebular SEDs, and commonly used sky subtraction strategies may interfere with the predicted contribution from the the nebula to the observed SED. While the relative flux contribution of the nebula -- and therefore the overall luminosity -- does depend on these effects, the {\it colours} of type A and type B galaxies are nonetheless likely to be similar as long as the system is young and nebular emission is dominant. The youth criterion comes in because the mismatch between the momentary stellar SED and the prior ionizing field to which the nebula responds becomes more severe once the high-mass stars of the stellar population have died and many intermediate-mass stars have evolved off the main sequence.

The nebular contribution from type A galaxies is modelled assuming a constant-density HII region with hydrogen density $n(\mathrm{H})=100$ cm$^{-3}$ and filling factor $f_\mathrm{fill}=0.01$. In the case of pop III galaxies, the gaseous metallicity is set to  $Z_\mathrm{gas}=10^{-7}$ (whereas $Z_\mathrm{stars}=0$), whereas pop II ($Z=0.0004$) and I ($Z=0.020$) galaxies have $Z_\mathrm{gas}=Z_\mathrm{stars}$. Scaled solar abundances are used in all cases. The default value for the gas covering factor of type A galaxies is $f_\mathrm{cov}=1$ (implying a Lyman continuum escape fraction $f_\mathrm{esc}=0$). However, in the case of a compact HII region, anisotropic feedback, supernova chimneys and irregularities in the gas density may result in large holes in the nebula, through which ionizing radiation can escape into the intergalactic medium. Simulations predict that the amount of Lyman continuum escape may be a function of dark halo mass \citep{Gnedin et al.,Razoumov & Sommer-Larsen,Yajima et al.}, but the likely escape fraction and its exact mass dependence remains controversial. In our model, the possibility of Lyman continuum escape through holes is treated by allowing the gas covering factor $f_\mathrm{cov}$ to take on values $0<f_\mathrm{cov}<1$. While having a very low but non-zero $f_\mathrm{cov}$ does not capture all the complexities of a type B galaxy, it still gives some indication of what to expect from such an object.

Type C objects are treated by setting $f_\mathrm{cov}=0$, which implies a Lyman-continuum escape fraction of unity and an SED composed of stars only. Since no nebula is produced, a photoionization code like Cloudy is not required to model this case. 

\section{The detection limits of high-redshift galaxies in JWST broadband surveys}
\label{masslimits}
\begin{figure*}
\plottwo{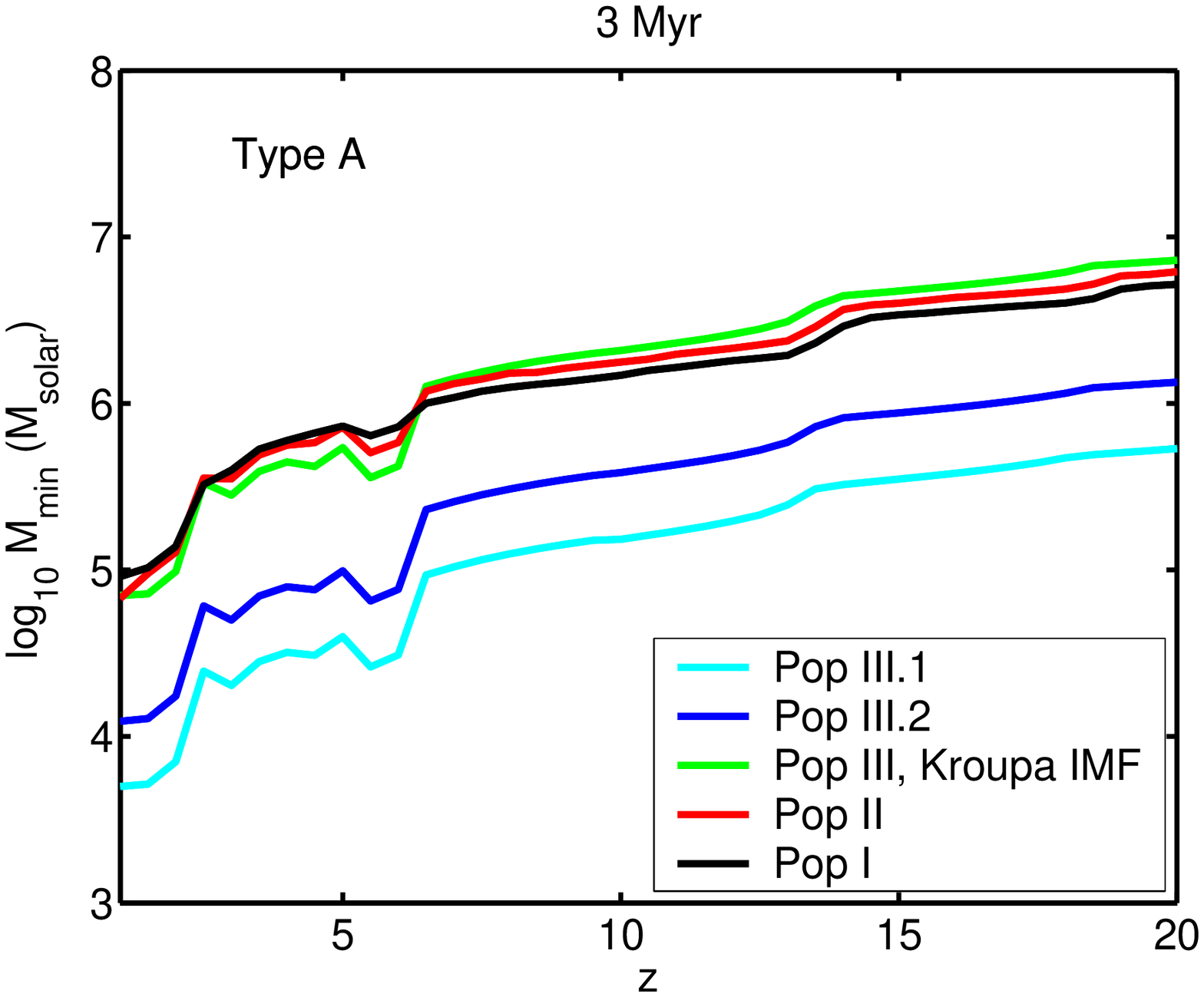}{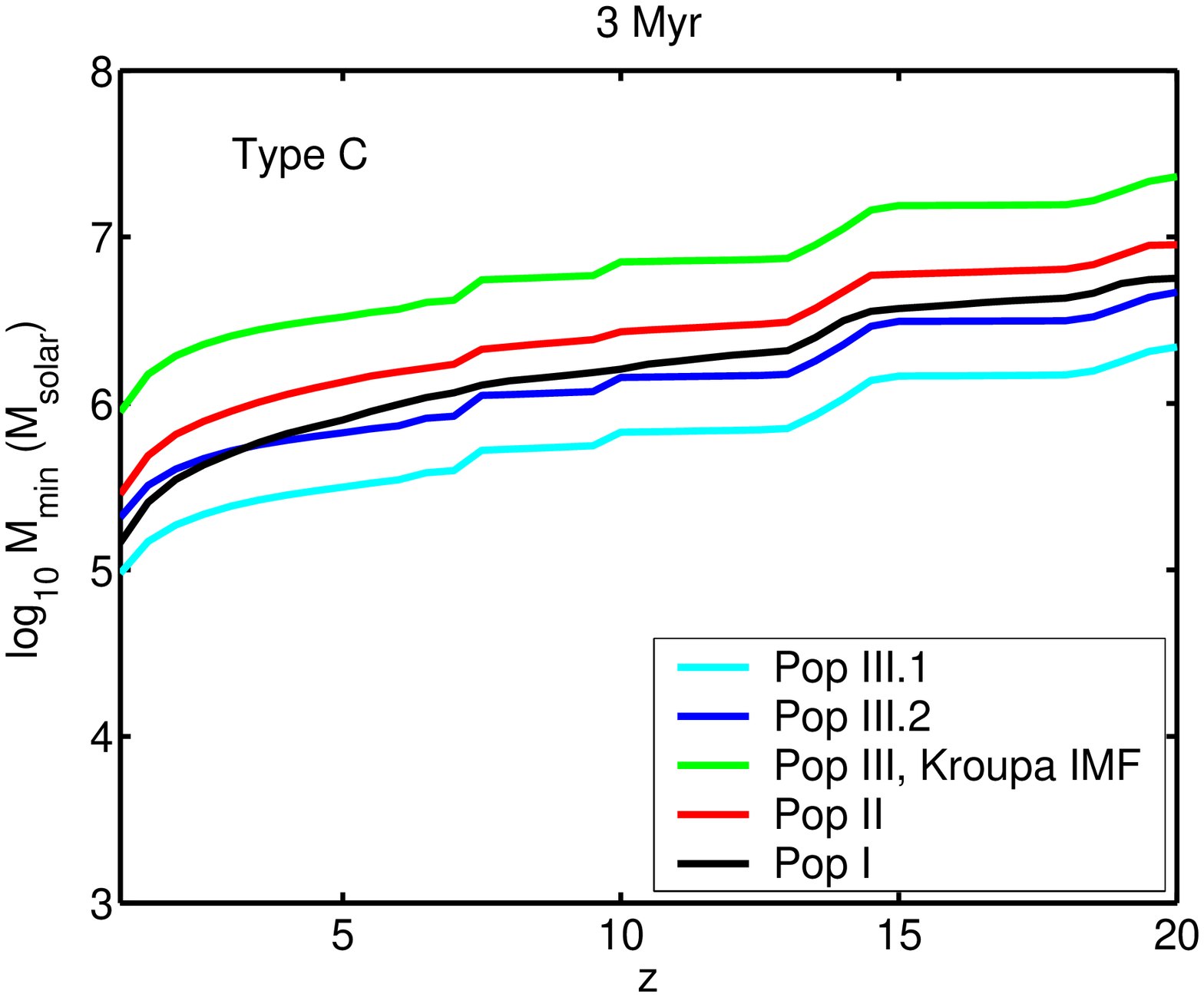}\\
\plottwo{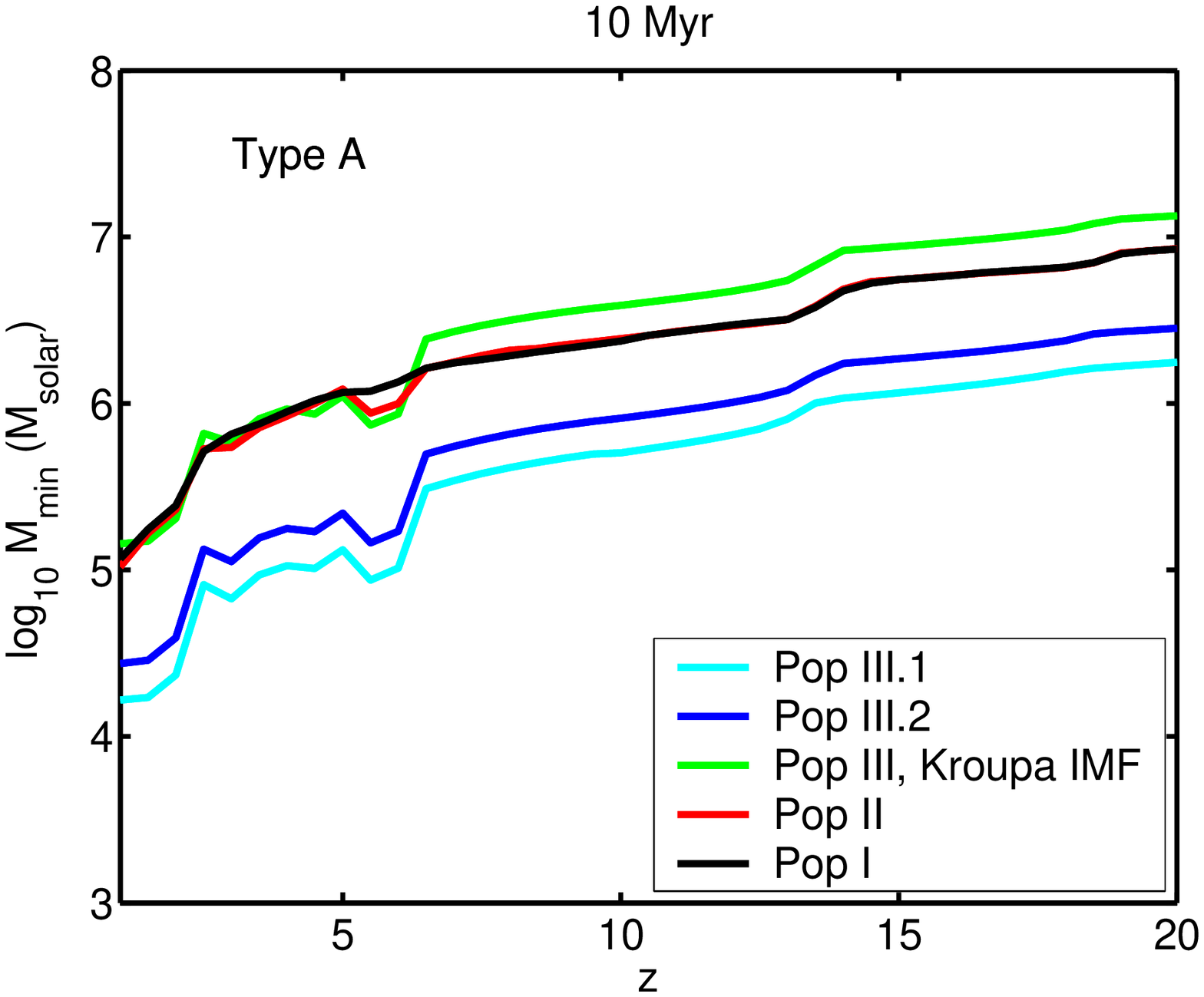}{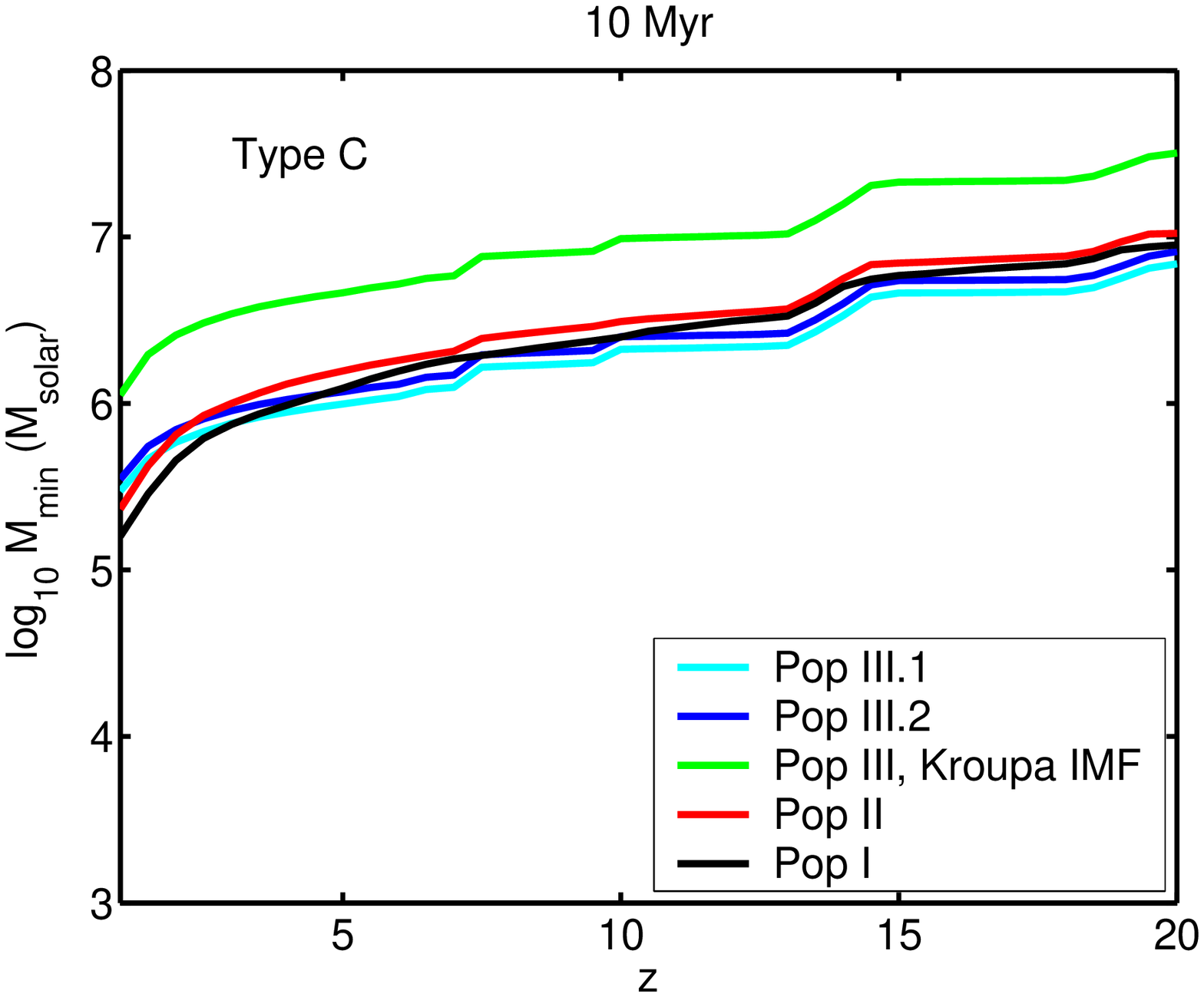}\\
\plottwo{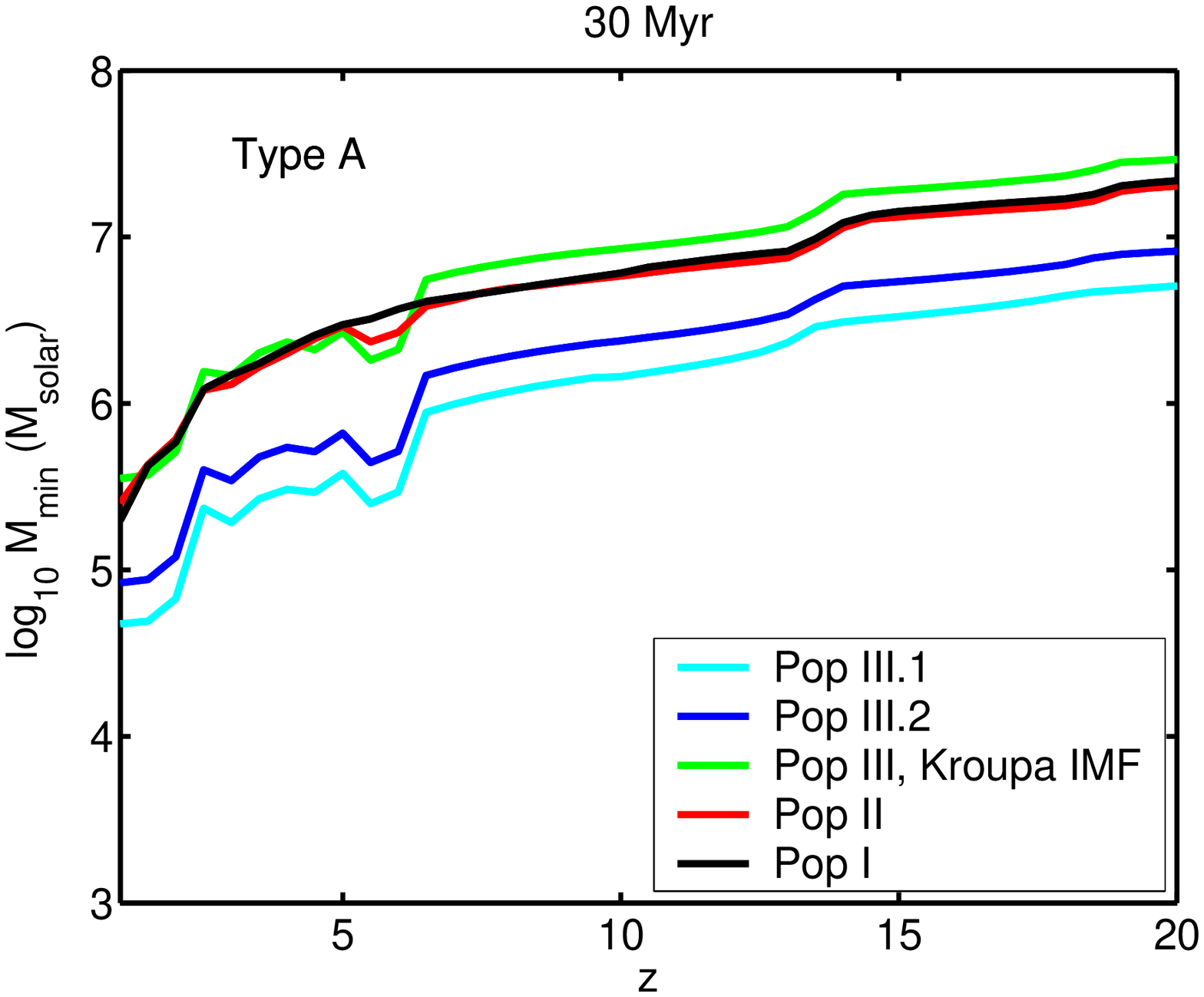}{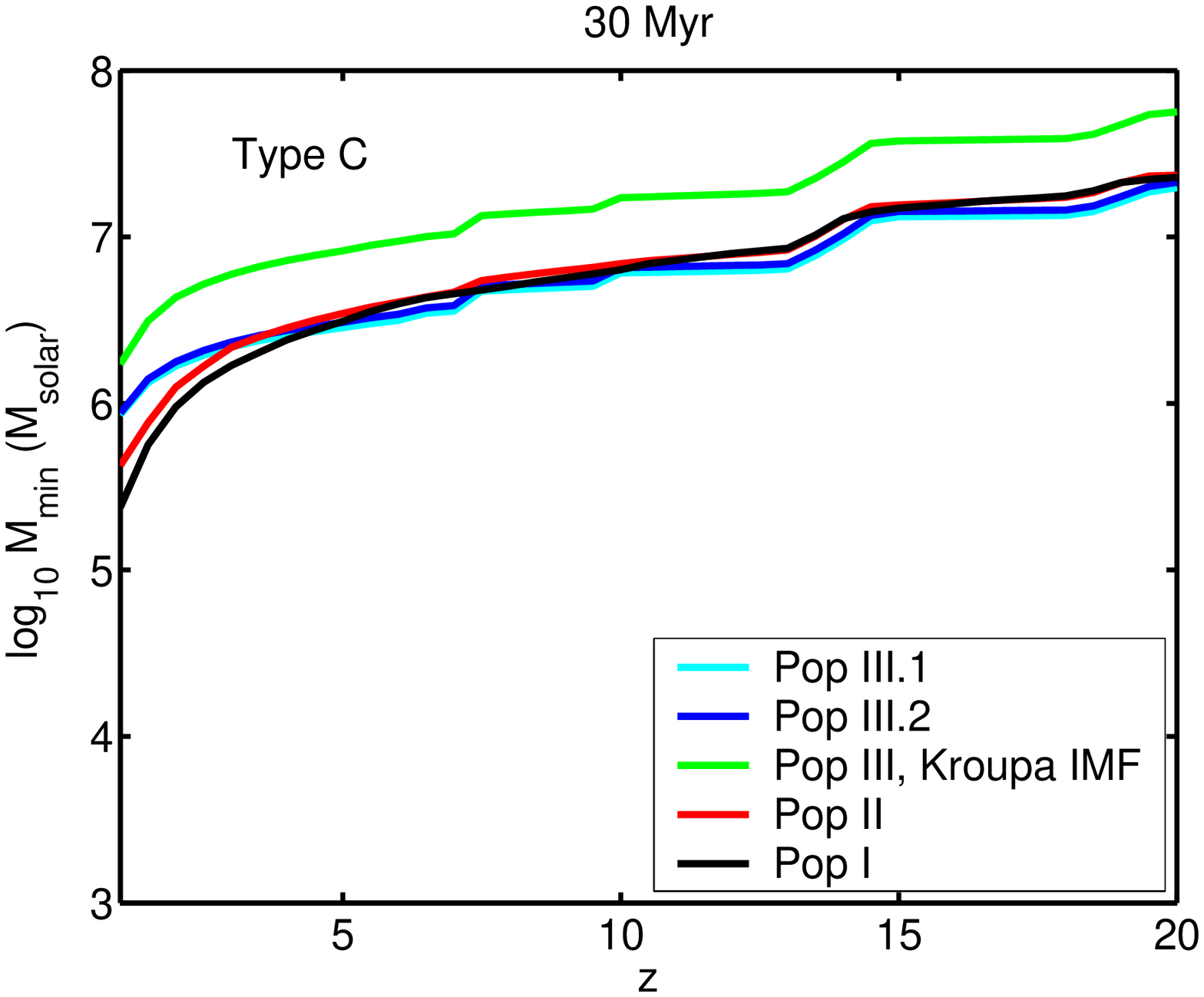}\\
\caption{Predicted mass detection limits in JWST ultra-deep fields. The panels show the lowest masses of burst-like stellar populations that JWST can detect through broadband imaging at 10$\sigma$ after a 100 h exposure, as a function of redshift. The line colours represent different combinations of population metallicities and IMFs: Pop III.1 (cyan line), pop III.2 (blue), pop III with a \citet{Kroupa} IMF (green), pop II (red line) and pop I (black line). The left column contains the predictions for type A galaxies (maximal nebular contribution) and the right column the corresponding predictions for type C galaxies (no nebular contribution). Each row corresponds to a different starburst age (3, 10 and 30 Myr from top to bottom) for a population undergoing a star formation episode with constant star formation rate.\label{Mmin}}
\end{figure*}

\begin{table}[t]
\begin{flushleft}
\caption[]{The $m_{AB}$ magnitiude limits adopted for 10$\sigma$ detections after $t_\mathrm{exp}=100$ h exposures, based on the NIRCam and MIRI sensitivities listed on the JWST homepage$^\mathrm{a}$.}
\begin{tabular}{llll} 
\tableline
Instrument & Filter & $\lambda$ ($\mu$m) & $m_{AB}$\cr
\tableline
NIRCam & F070W  & 0.70 & 29.9\cr
& F090W & 0.90 & 30.4\cr
& F115W & 1.15 & 30.6\cr
& F150W  & 1.50 & 30.7\cr
& F200W  & 2.00 & 30.9\cr
& F277W  & 2.77 & 30.6\cr
& F356W  & 3.56 & 30.6\cr
& F444W  & 4.44 & 30.1\cr
MIRI & F560W  & 5.60 & 27.6\cr
& F770W  & 7.70 & 27.3\cr
& F1000W  & 10.0 & 26.3\cr
& F1130W  & 11.3 & 25.3\cr
& F1280W  & 12.8 & 25.5\cr
& F1500W  & 15.0 & 25.2\cr
& F1800W  & 18.0 & 24.3\cr
& F2100W  & 21.0 & 23.5\cr
& F2550W  & 25.5 & 22.3\cr
\tableline
\tableline
\end{tabular}\\
$^\mathrm{a}$ http://www.stsci.edu/jwst/instruments/
\label{ABmag_limits}
\end{flushleft}
\end{table}
In Fig.~\ref{Mmin}, we use Yggdrasil to predict the population masses of the faintest star-forming galaxies detectable through JWST broadband imaging when taking extremely long exposures of a small patch of the sky, i.e. a so-called Ultra Deep Field (UDF). The mass limits presented are based on the requirement that galaxies are detected at 10$\sigma$ in {\it at least} one JWST broadband filter (spectral resolution $R=4$) after a 100 h exposure (per filter). While the detailed specifications of JWST UDFs have not yet been fixed, 100 h represents a reasonable estimate of the longest exposure times that are likely to be relevant. The resulting mass limits can easily be rescaled to other exposure times $t_\mathrm{exp}$ using:
\begin{equation} 
M_\mathrm{min}\propto \left(\frac{t_\mathrm{exp}}{100\; \mathrm{h}} \right)^{1/2}.
\end{equation}
Our derived mass limits take {\it all} of the 17 JWST broadband ($R=4$) filters into account, and are based on the actual JWST transmission curves of these filters\footnote{The one exception is the NIRCam/F090W filter profile, which was not available to us at the time of writing. In this case, the results are instead based on a top-hat transmission curve.} and the estimated AB-magnitude limits listed in Table.~\ref{ABmag_limits}. These 10$\sigma$, $t_\mathrm{exp}=100$ h detection limits $m_\mathrm{AB,\; 100 h}$ can be rescaled to other exposure times using: 
\begin{equation} 
m_\mathrm{AB}=2.5\log_{10}\left[\left(\frac{t_\mathrm{exp}}{100\; \mathrm{h}}\right)^{1/2}\right] + m_\mathrm{AB,\; 100 h}.
\end{equation}

\begin{figure*}
\plottwo{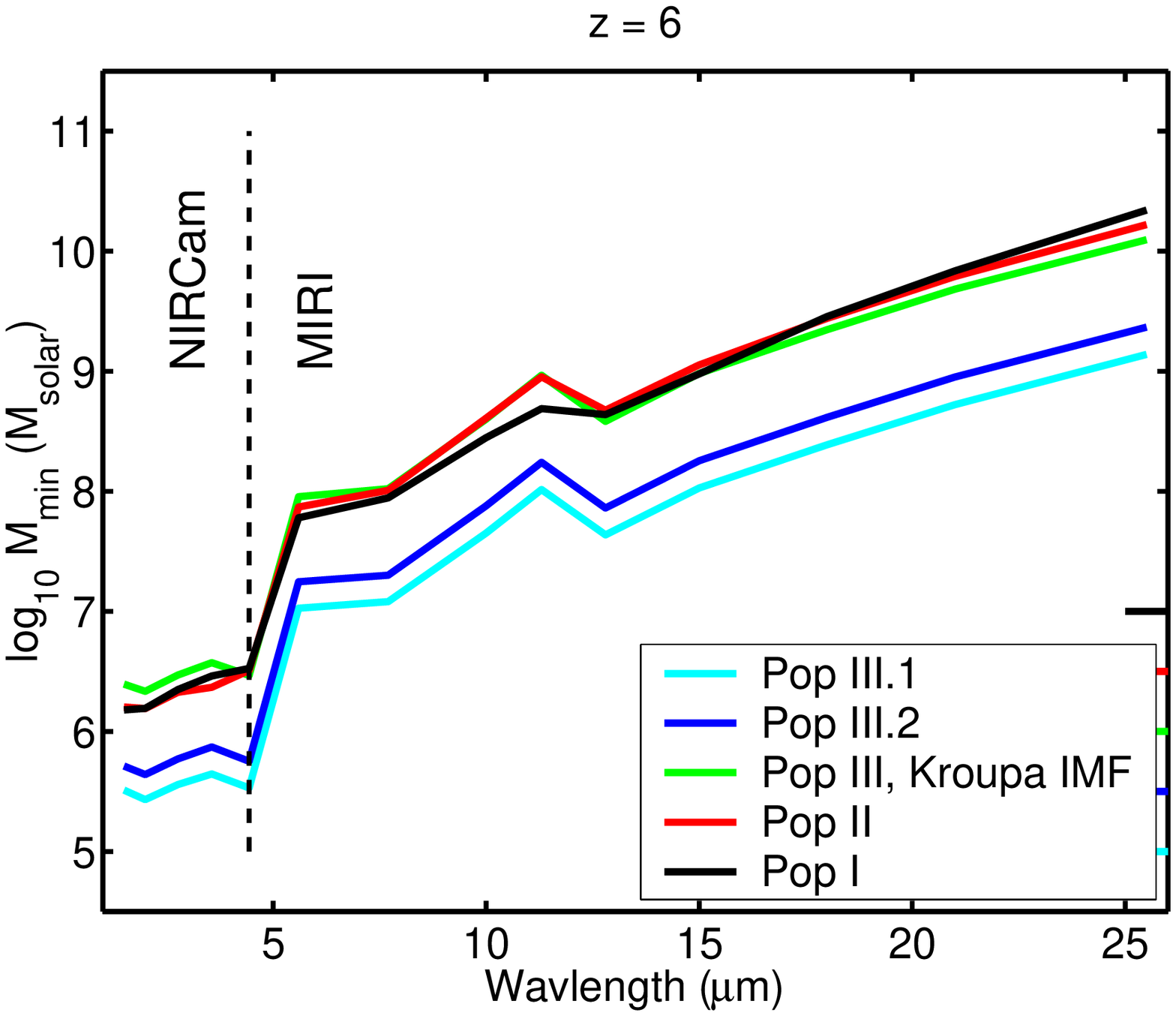}{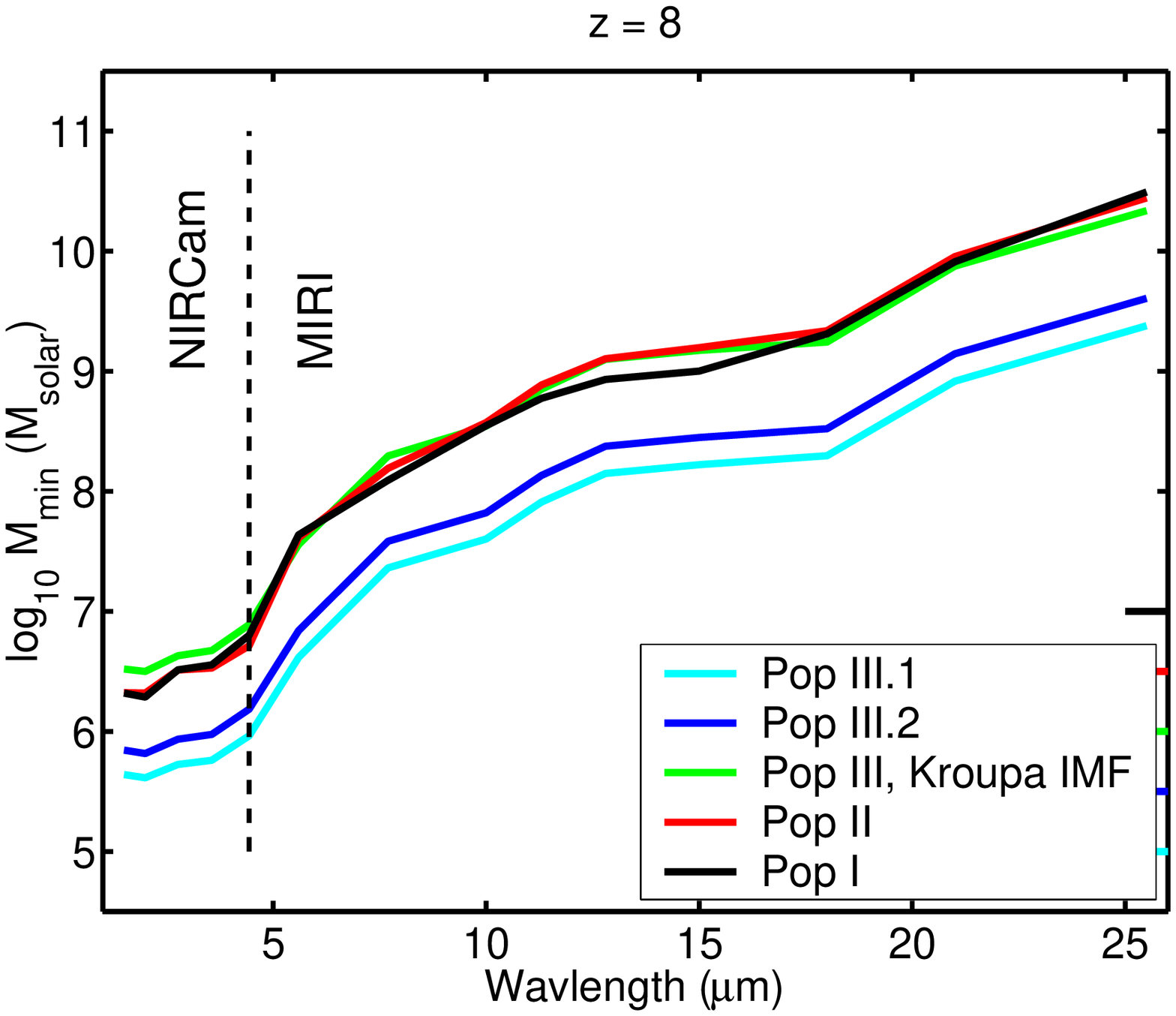}
\plottwo{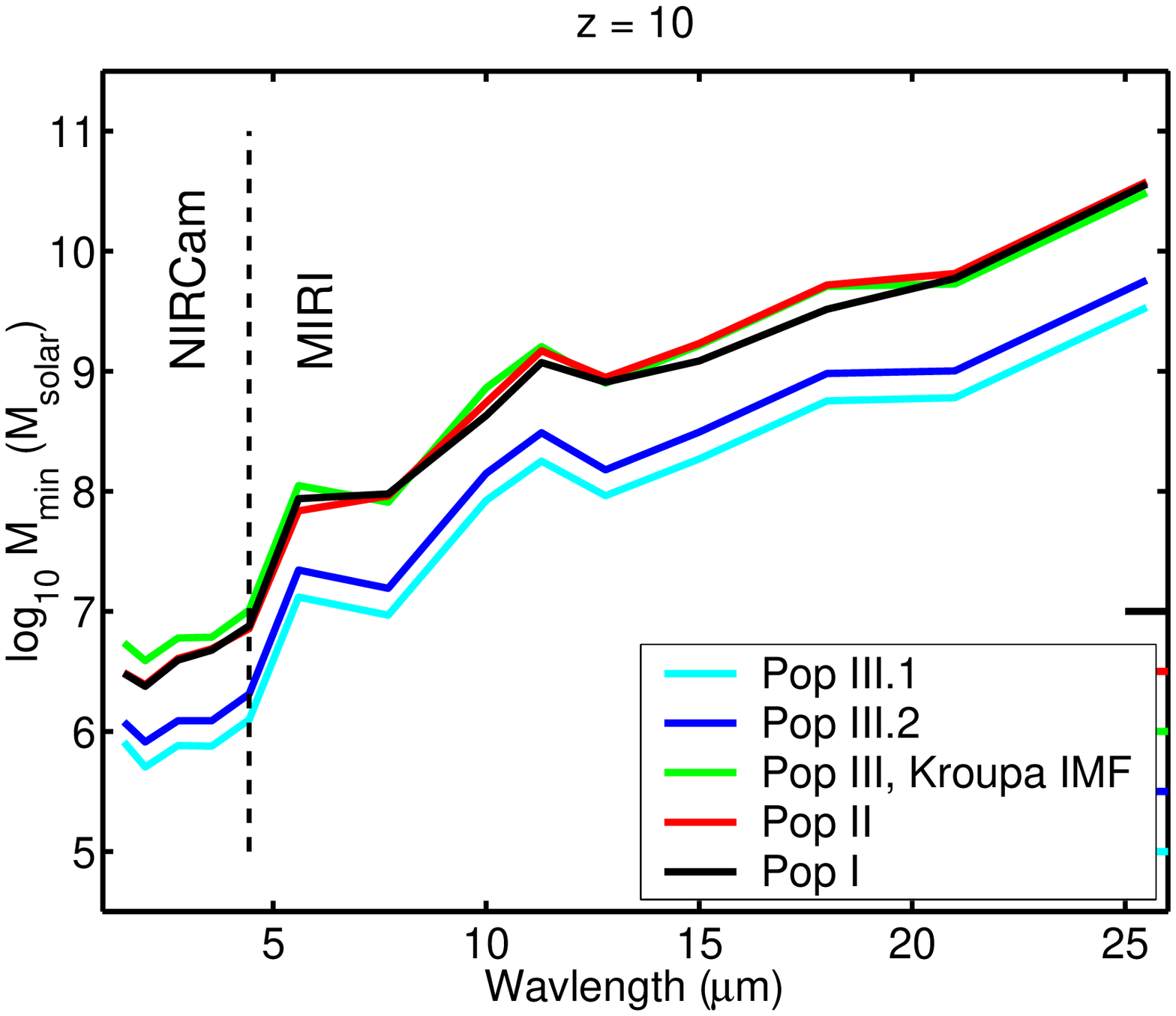}{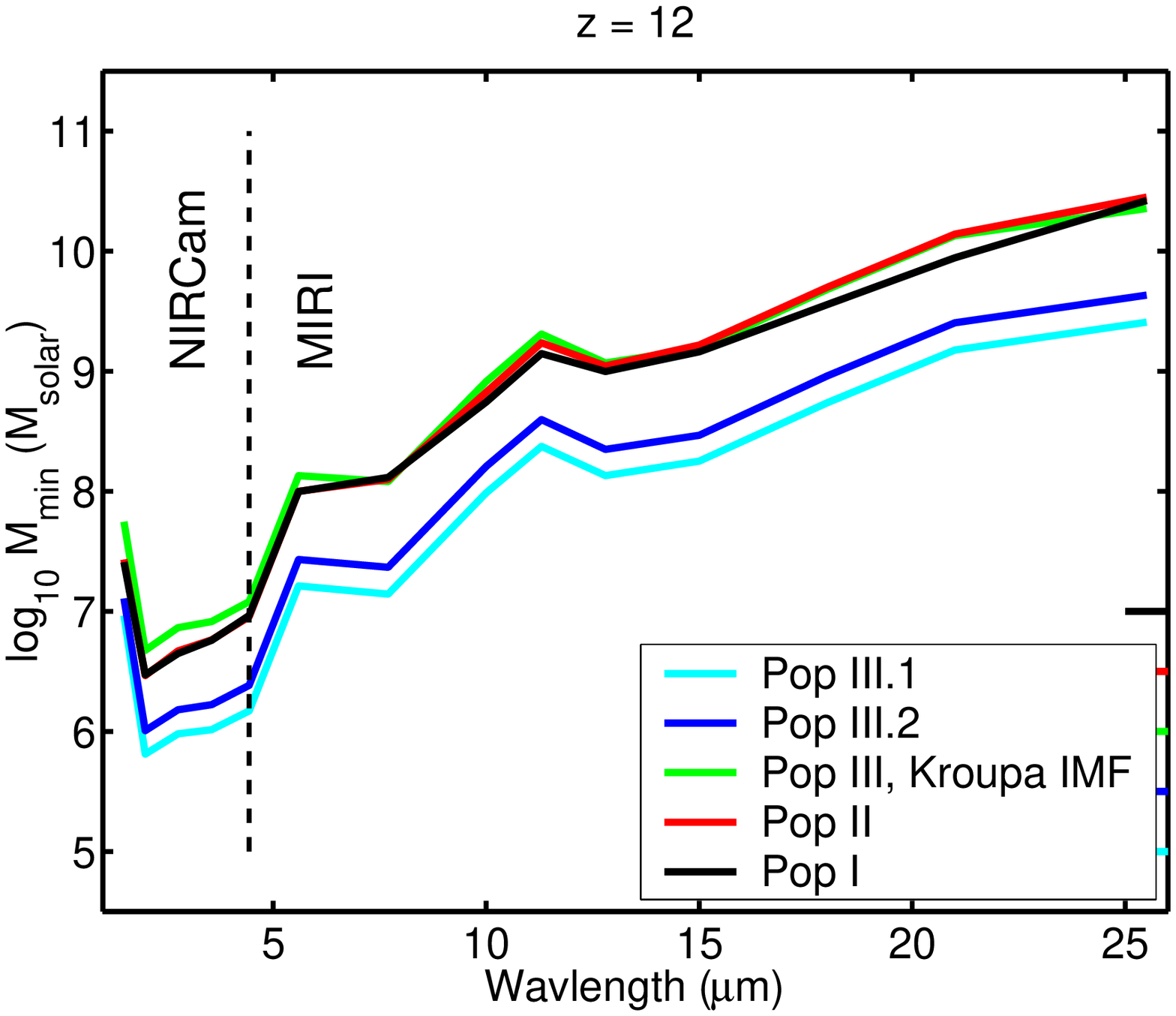}
\caption{Predicted mass detection limits for young, type A galaxies in JWST ultra-deep fields as a function of filter wavelength. The panels show the lowest stellar population masses of burst-like, 10 Myr old, type A (maximal nebular contribution) galaxies that JWST can detect through broadband imaging at 10$\sigma$ after 100 h exposures per filter, as a function of central filter wavelength. The line colours have the same meaning as in Fig.~\ref{Mmin}. Each panel corresponds to a different redshift ($z=6$, 8, 10, 12). No results are included for filters at $\leq 1.15\ \mu$m, since IGM absorption is likely preventing detection in these filters at some of these redshifts. The sharp increase in the lowest detectable mass for the F150W filter (at 1.50 $\mu$m) at $z=12$ is also due to this effect. The dashed vertical line in each panel delimits the wavelength ranges of the NIRCam to MIRI instruments.  
\label{Mmin_singlez}}
\end{figure*}

To avoid predictions hinging on the highly uncertain luminosity of the Ly$\alpha$ emission line at high redshifts \citep[e.g.][]{Fontana et al.,Hayes et al.}, we assume the Ly$\alpha$ escape fraction to be zero -- i.e. the Ly$\alpha$ line does not contribute to the predicted fluxes at all. When computing the JWST broadband fluxes, we furthermore set all SED fluxes shortward of Ly$\alpha$ to zero for model galaxies at $z>6$, to reflect the high level of absorption in the neutral intergalactic medium at these epochs.

The mass limits in Fig.~\ref{Mmin} are presented for pop I, pop II, pop III galaxies of both type A (maximal contribution from the nebula) and C (no contribution from the nebula) with constant star formation rates as a function of age (up to 30 Myr) and redshift (up to $z=20$). As discussed in Sect.~\ref{scenarios}, galaxies starting from zero metallicity are not likely to retain their pop III characteristics for much longer than this. While pop I and pop II galaxies may well have higher ages at these epochs, such objects would need to be even more massive than suggested by the limits at 30 Myr in Fig.~\ref{Mmin} to be detectable with JWST. Hence, Fig.~\ref{Mmin} places hard limits on what can realistically be observed with this telescope through ultra deep imaging in the absence of gravitational lensing.
 
The mass used to define these limits is the mass of gas converted into stars since the beginning of a star formation episode:
\begin{equation}
M(t)=\int_0^t \mathrm{SFR}(t')\; \mathrm{d}t'.
\label{Mass_eq}
\end{equation}
This is equivalent to the population mass often discussed in the context of simulations, where the overall gas mass is multiplied by the star formation efficiency of the first starburst episode to compute the gas converted into stars. The mass in {\it luminous stars} (stars that have not yet exploded as supernovae or turned into compact remnants) at these ages can be considerably lower at ages $\gtrsim 3$ Myr (especially in the case of top-heavy IMFs), since many of the stars forming at $t=0$ yr have then already faded away. For instance, the mass in luminous stars in a pop III.1 galaxy with constant star formation is lower than the plotted limits by a factor of $\approx 5$ after 10 Myr, and by a factor of $\approx 14$ at 30 Myr. On the other hand, the mass in luminous stars within \citet{Kroupa} IMF populations is reduced by only $\approx 20$ \% (pop III) and $\approx 10$ \% (pop II and I) after 30 Myr of constant star formation. In the case of star formation episodes with finite duration $\tau$, $M(t)$ equals $M_\mathrm{tot}$ defined in eq.(\ref{Mtot_eq}) if evaluated at a time after the end of the starburst ($t\geq \tau$).

As seen in Fig.~\ref{Mmin}, $M\sim10^6\ M_\odot$ star-forming pop I and pop II galaxies (black and red lines) can be detected at $z\approx 10$, whereas pop III.1 galaxies (cyan lines) can be detected even if the mass is as low as $M\sim10^5\ M_\odot$. Pop III.2 galaxies (blue lines) lie between pop I/II and pop III.1 galaxies in this diagram. On the other hand, pop III galaxies with a \citet{Kroupa} IMF (green lines) tend to require a mass at least as high as pop I/II galaxies (and sometimes significantly higher) to be detected, because of their lower stellar mass-to-light ratios in these passbands.

For galaxies with a given IMF and metallicity, Type A objects (maximal nebular flux contribution) are always brighter than type C objects (no nebular flux contribution) and can therefore be detected at lower masses. However, the difference is much smaller for pop I and II galaxies (less than a factor of 2 at $z=10$) than for pop IIIs, due to the higher relative ionizing fluxes of the latter objects. For instance, nebular emission allows young ($\approx 3$ Myr old) pop III.1 galaxies at $z=10$ to be detected at masses a factor of $\approx 4$ lower than in the case where all the ionizing radiation is escaping into the intergalactic medium. The mass detection limits presented for pop III and pop II galaxies in Fig.~\ref{Mmin} are similar (within a factor of 2, once differences in exposure times are considered) to those derived by \citet{Pawlik et al.}, who considered the JWST detectability of lines and continuum radiation separately.

The detection limits in Fig.~\ref{Mmin} indicate the masses of galaxies that JWST would be able to detect at a single wavelength, but to derive the properties of the objects observed, photometry in several filters is always required. Since the JWST sensitivity varies greatly as a function of wavelength, the detection limits are sensitive to the filters used. In Fig.~\ref{Mmin_singlez}, we plot the lowest stellar population masses of 3, 10 and 30 Myr old type A galaxies (maximal nebular emission) detectable in a JWST UDF (10$\sigma$ detections after 100 h exposures) as a function of central filter wavelength at $z=6$, 8, 10 and 12. The mass limits stay approximately constant within the wavelength range of the JWST/NIRCam instrument (operating at 0.7--4.4 $\mu$m), but then rise by an order of magnitude as one switches to JWST/MIRI observations at $\geq 5.5\ \mu$m and continue to deteriorate almost monotonically all the way up to the 25 $\mu$m limit of MIRI. Hence, detections of high-redshift galaxies in the MIRI filters will be very expensive in terms of observing time, compared to those in NIRCam filters. On average, the relative contribution from nebular continuum grows with increasing rest-frame wavelength, and this compensates for some of the sensitivity loss in the MIRI bands. For galaxies without nebular emission (type C), the decline in mass sensitivity with increasing JWST passband wavelength is even more dramatic than plotted here.  

\section{The spectral signatures of type A, pop III galaxies}
\label{typeA}
As discussed in Sect.~\ref{intro}, all searches for pop III galaxies with largely nebular SEDs (our type A) have so far focused on the fluxes of the Ly$\alpha$ line, the HeII $\lambda1640$, $\lambda4686$ lines, or the Lyman `bump' \citep[e.g.][]{Tumlinson & Shull,Tumlinson et al. a,Oh et al.,Malhotra & Rhoads,Schaerer a,Schaerer b,Inoue}. However, neither of these spectral features is likely to produce a robust and detectable signature in broadband filters for objects at $z>6$. HeII is too weak, the Lyman `bump' is absorbed by the intergalactic medium and the escape fraction of  Ly$\alpha$ photons is notoriously difficult to predict. 

Here, we propose a different strategy, based on the fact that such galaxies are likely to have strong hydrogen and helium emission lines, but -- due to the lack of metals -- no lines due to heavier elements. Certain rest frame UV and optical metal emission lines, most notably [OII]$\lambda$3727, [OIII]$\lambda$5007 and [SIII]$\lambda$9069, are sufficiently strong to influence the broadband fluxes of galaxies \citep[e.g.][]{Anders & Fritze-v. Alvensleben,Zackrisson et al. b,Schaerer & de Barros a}, and removing these lines will therefore alter the broadband colours of type A objects. Recently, \citet{Inoue b} argued that the lack of [OIII]$\lambda$5007 emission would be one of the best ways to single out pop III galaxies. Our analysis agrees with this conclusion. 

\subsection{NIRCam colour criteria}
\begin{figure}
\plotone{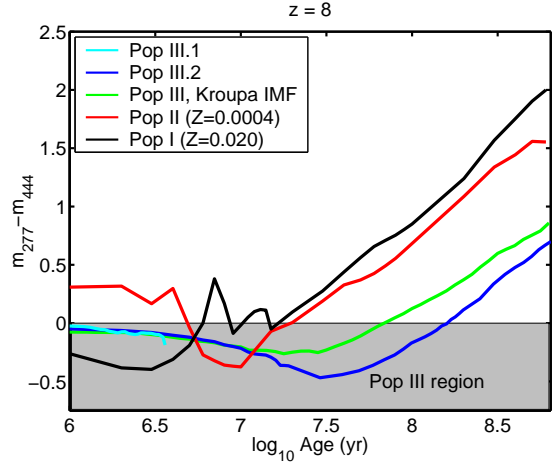}
\caption{The $m_{277}-m_{444}$ colours of type A, instantaneous-burst galaxies at $z=8$. The line colours indicate different metallicity and IMF combinations: Pop III.1 (cyan), Pop III.2 (blue), Pop III with a \citet{Kroupa} IMF (green), Pop II (red) and Pop I (black). A \citet{Kroupa} IMF has been assumed for all of the $Z=0.0004$--0.020 populations. \citet{Inoue b} suggests that $m_{277}-m_{444}<0$ (shaded region) should serve as a colour criterion for pop III galaxies, but as seen, both Pop II (red line) and Pop I (black solid line) objects can meet this colour criterion during certain phases of their evolution. While the details of the IMF determines the longevity of pop III galaxies, the $m_{277}-m_{444}$ colours predicted for objects of age $\lesssim 10$ Myr are {\it not} sensitive to the exact pop III IMF, as illustrated by the overlapping cyan, blue and green lines. The age axis runs up to the age of the Universe at this redshift ($\approx 6.5\times 10^8$ yr).
\label{typeA_Inoue_crit}}
\end{figure}

\citet{Inoue b} suggests that imaging in the two JWST/NIRCam filters F277W and F444W could be useful to single out pop III galaxies at $z\approx 8$, since the lack of [OIII] emission in the F444W filter would render the colours of such objects very blue. However, the $m_{277}-m_{444}\leq 0$ criterion proposed by Inoue for pop III objects is not unique, as demonstrated using instantaneous-burst models in Fig.~\ref{typeA_Inoue_crit} -- both pop II and pop I galaxies of type A can at certain ages venture blueward of this limit. In the case of our pop II model (red line), this happens because of a short-lived evolutionary phase at ages of $\approx 10-20$ Myr, where the ionizing flux has dropped sufficiently to make the nebular contribution irrelevant, yet the UV continuum of the stars remains very blue. In the case of our pop I model ($Z=0.020$; black solid line), this happens because very young ($\lesssim 5$ Myr) populations with high metallicities also display very weak [OIII] emission lines \citep[e.g.][]{Panagia,Nagao et al. b}. In an instantaneous burst scenario, the timescales over which which Pop I/II systems display these very blue colours are shorter than the corresponding timescales for pop III systems, which may limit the risk of pop I/II interlopers in surveys for pop III galaxies. On the other hand, the typical star formation histories may be different in the two cases, and adopting a more extended star formation history for pop I/II could make these objects display pop III-like colours for a longer period of time. Dust extinction, which is likely to be associated with star formation in high-metallicity environments, may of course render the colours of young pop I/II systems redder and possibly prevent metal-enriched galaxies from entering colour-selected pop III samples.

While the details of the IMF determines the longevity of pop III galaxies, the $m_{277}-m_{444}$ colours predicted for type A, pop III objects of age $\lesssim 10$ Myr are not sensitive to the exact pop III IMF, as illustrated by the largely overlapping cyan, blue and green lines in Fig.~\ref{typeA_Inoue_crit}.

\subsection{Combined NIRCam and MIRI colour criteria}
\begin{figure*}
\plottwo{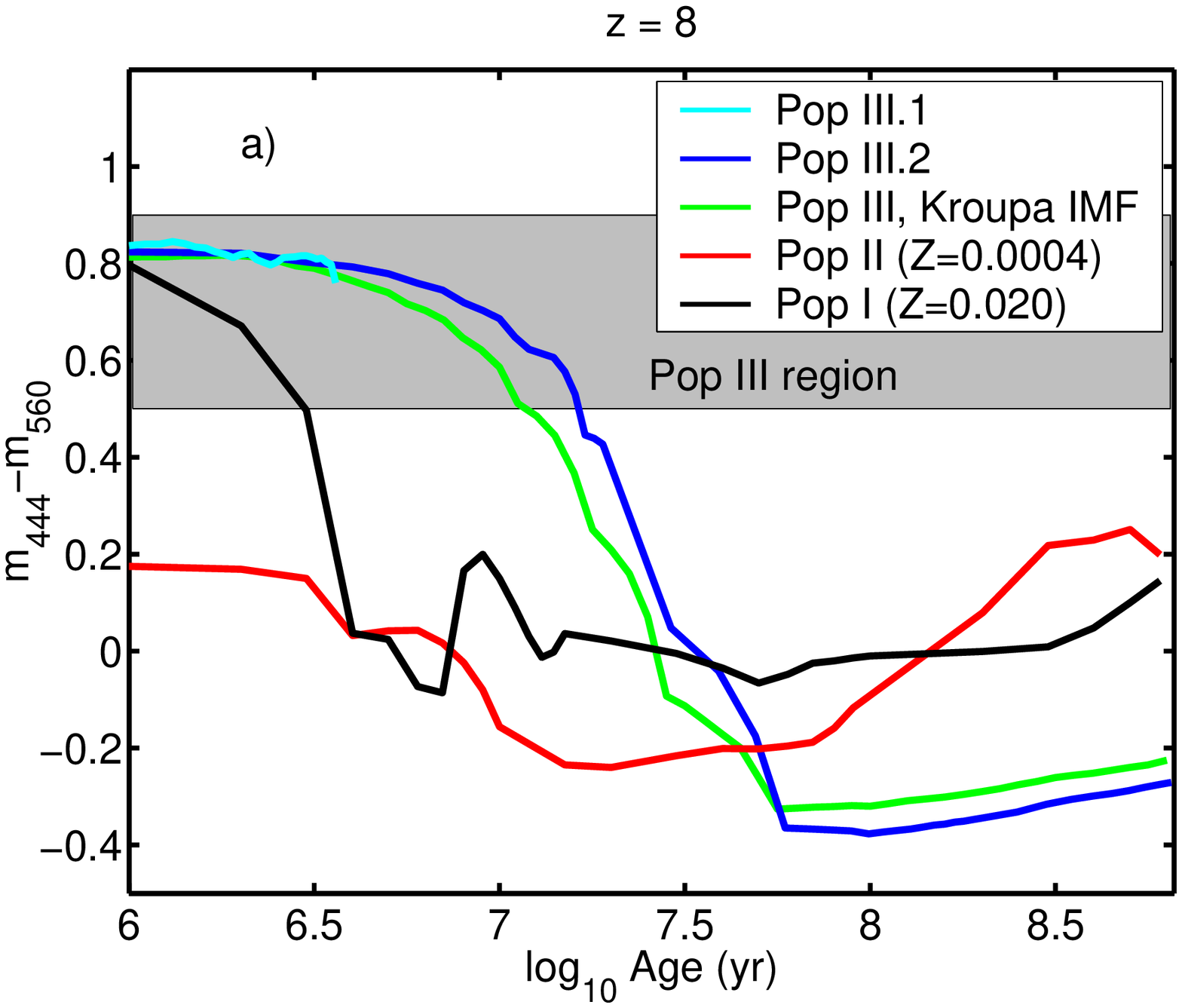}{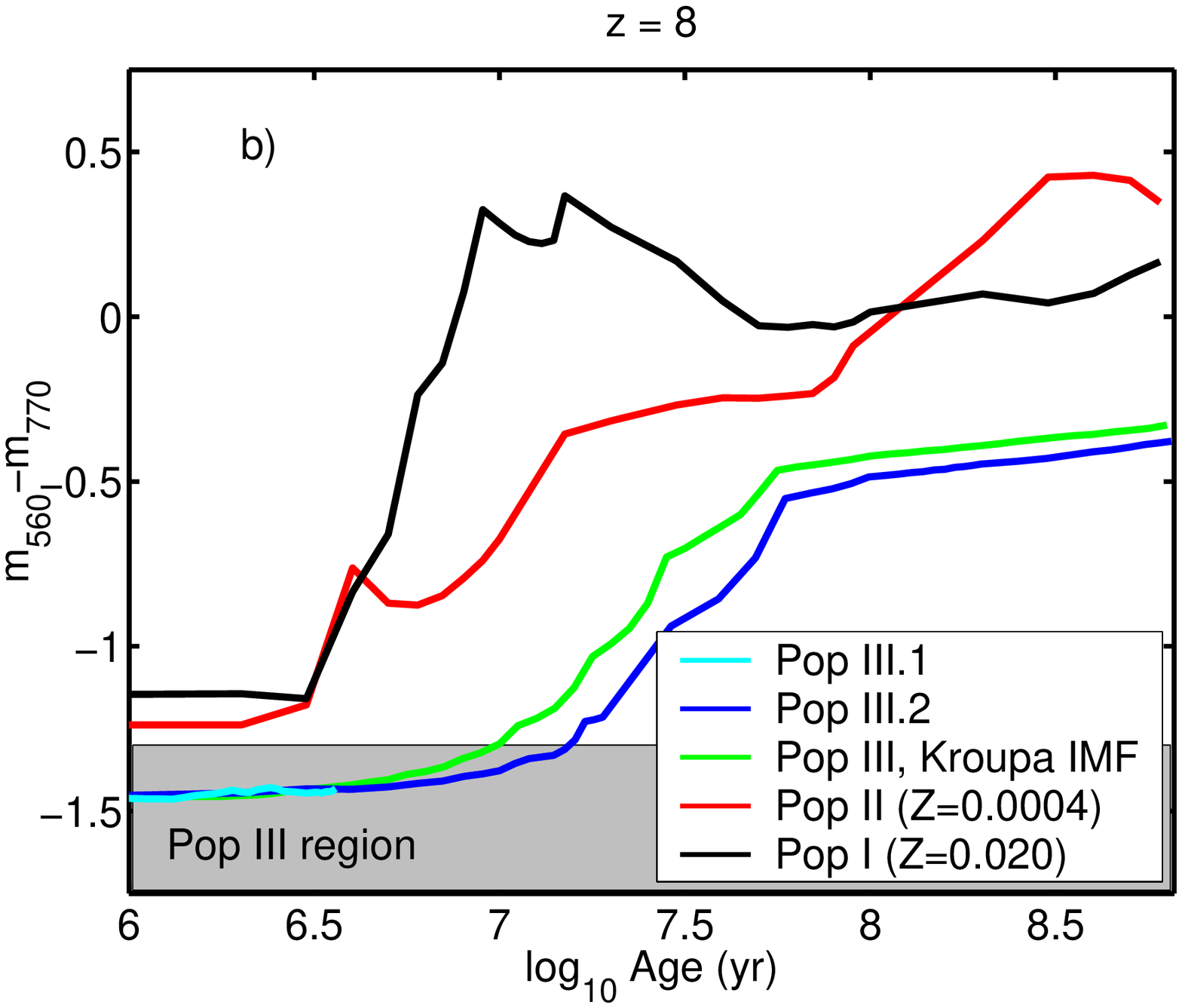}
\caption{The $m_{444}-m_{560}$ ({\bf a}) and $m_{560}-m_{770}$ ({\bf b}) colours of type A, instantaneous-burst galaxies at $z=8$. The line colours represent different metallicity and IMF combinations: Pop III.1 (cyan), Pop III.2 (blue), Pop III with a \citet{Kroupa} IMF (green), Pop II (red) and Pop I (black). A \citet{Kroupa} IMF has been assumed for the two latter populations. In $m_{444}-m_{560}$, young ($\lesssim 15$ Myr old) pop III galaxies appear redder than galaxies of almost all other ages and metallicities due to the absence of strong [OIII] emission in the $m_{444}$ filter, yet strong H$\alpha$ emission in $m_{560}$. The only notable exceptions are newborn ($\lesssim 3$ Myr old) pop I galaxies, which also display very red colours. In $m_{560}-m_{770}$, pop III galaxies appear bluer than all other objects at this redshift due to their high H$\alpha$ equivalent widths. While the details of the IMF determines the longevity of pop III galaxies, the colours predicted during their lifetimes are {\it not} sensitive to the exact pop III IMF, as illustrated by the largely overlapping cyan, blue and green lines. Shaded regions indicate the regions where pop III galaxies would be identifiable. In both panels, the age axis runs up to the age of the Universe at this redshift ($\approx 6.5\times 10^8$ yr). The error bars on the observed colours would be $\approx 0.15$ mag in the case of $10\sigma$ detections in the relevant filters.  
\label{typeA_MIRIcrit}}
\end{figure*}

\begin{figure}
\plotone{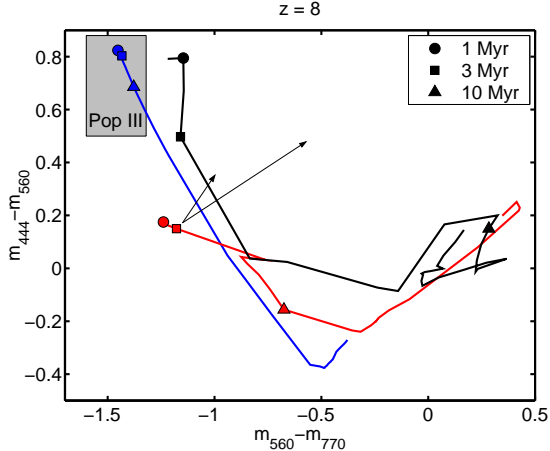}
\caption{The signatures of type A, pop III galaxies in $m_{560}-m_{770}$ vs. $m_{444}-m_{560}$ diagram at $z=8$. The lines correspond to a limited set of the instantaneous-burst models presented in Fig.~\ref{typeA_MIRIcrit} (to avoid cluttering): Pop III.2 (blue line), Pop II (red line) and Pop I (black line). Markers along the tracks indicate ages of 1 Myr (circle), 3 Myr (square) and 10 Myr (triangle). The tracks then continue up to the age of the Universe at this redshift ($\approx 6.5\times 10^8$ yr). Arrows indicate how the colours of a 3 Myr old pop II galaxy would be shifted in the case of LMC extinction (short arrow) and \citet{Calzetti} extinction (long arrow) for rest frame $E(B-V)=0.25$ mag. In the case of Calzetti extinction, this refers to the extinction affecting by the stellar component SED, whereas that affecting the nebular component is higher by a factor $\approx 2.3$. The reddening vectors are similiar (but not identical) for other ages and metallcities. In the absence of extinction, young pop III galaxies occupy a unique corner (shaded region) of this diagram. While extinction (if at all effective in pop III objects) could potentially shift pop III galaxies into regions occupied by young pop I objects, dust cannot make other types of galaxies mimic the intrinsic colours of pop III objects. This diagnostic diagram looks approximately the same throughout the redhsift range $z=7$--8, as discussed in the main text. The error bars on the observed colours would be $\approx 0.15$ mag in the case of $10\sigma$ detections in the relevant filters. 
\label{typeA_MIRIcolcol}}
\end{figure}

In Fig.~\ref{typeA_MIRIcrit}, we demonstrate that that cleaner selection criteria for type A, pop III galaxy candidates at $z\approx8$ can in principle be derived by combining JWST/NIRCam data in the F444W filter with observations in two JWST/MIRI filters: F560W and F770W. The lack of [OIII]$\lambda$5007 in F444W and the presence of H$\alpha$ in F560W makes young pop III galaxies redder in $m_{444}-m_{560}$ (Fig.~\ref{typeA_MIRIcrit}a), yet bluer in $m_{560}-m_{770}$ (Fig.~\ref{typeA_MIRIcrit}b) than other types of galaxies. Due to the [OIII] degeneracy between pop III and pop I galaxies, both can attain similar $m_{444}-m_{560}$ colours. However, since there is no similar degeneracy in $m_{560}-m_{770}$, pop III objects appear in a corner of the $m_{560}-m_{770}$ vs. $m_{444}-m_{560}$ diagram (Fig.~\ref{typeA_MIRIcolcol}) where no other galaxies should appear. Pop III objects remain in the upper left corner (grey region) up to $\approx 15$ Myr after formation in the case of an instantaneous burst, and longer in the case of more extended star formation histories (in principle up to $\sim 100$ Myr in the case of a constant SFR, although such high ages are unlikely to apply, as discussed in Sect.~\ref{scenarios}). The colours of pop III galaxies with different IMFs are almost identical at young ages. We have verified that these criteria hold for objects up to an age of $\approx 15$ Myr as long as the Lyman-continuum escape fraction is $f_\mathrm{esc}\lesssim 0.5$, irrespective of the pop III IMF (but higher $f_\mathrm{esc}$ are allowed in the case of younger ages or more top-heavy IMFs). Fig.~\ref{typeA_MIRIcolcol} also illustrates the effect of dust reddening on these colour criteria. Even though substantial dust reddening (if at all possible in pop III galaxies) could conceal pop III galaxies by making them appear as newborn, high-metallicity objects, there is no risk of false positives - i.e. no other type of galaxies in our model grid can spuriously appear as pop III objects at this redshift. 

\subsection{The redshift evolution of type A galaxy colours}
\begin{figure*}
\plottwo{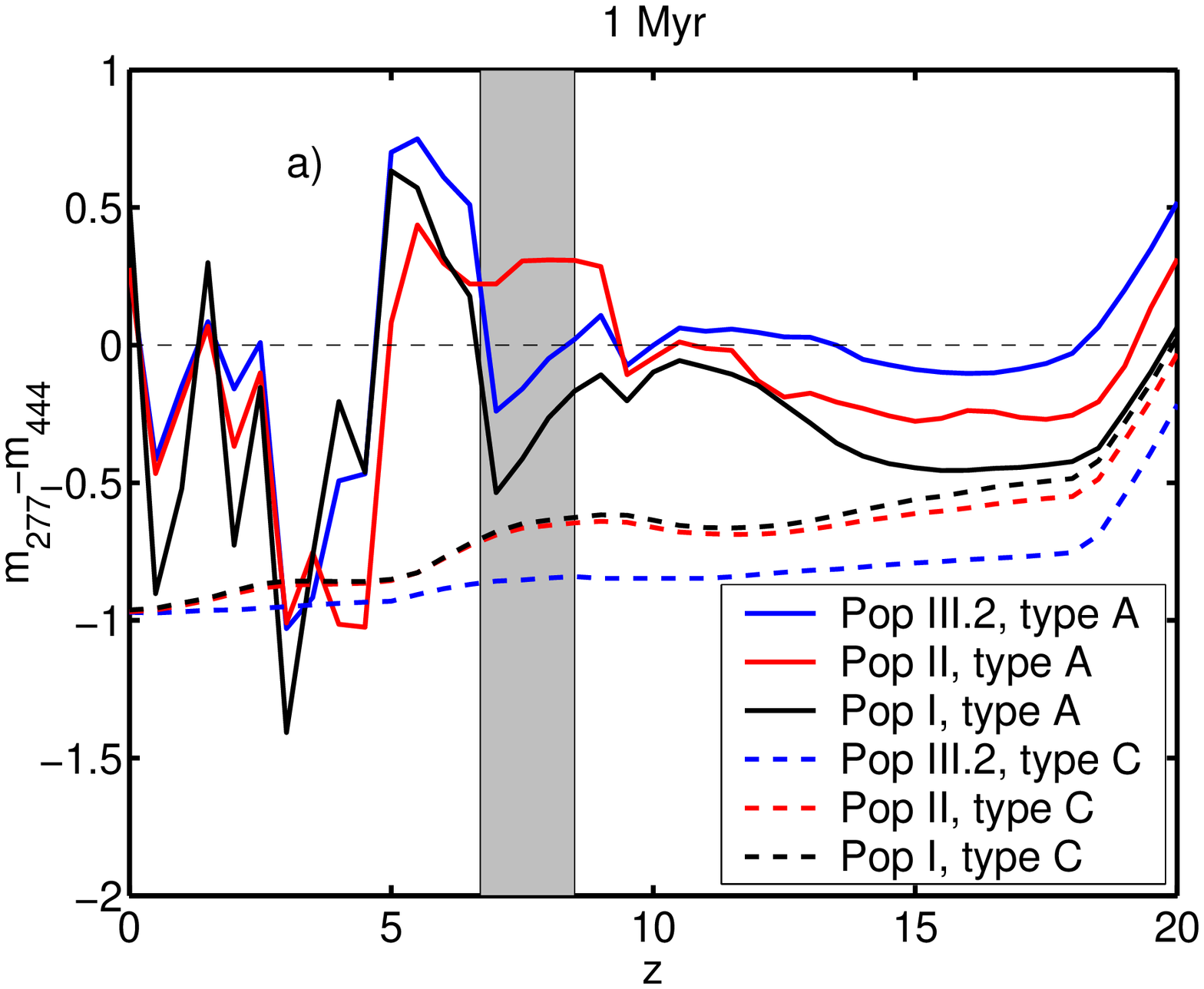}{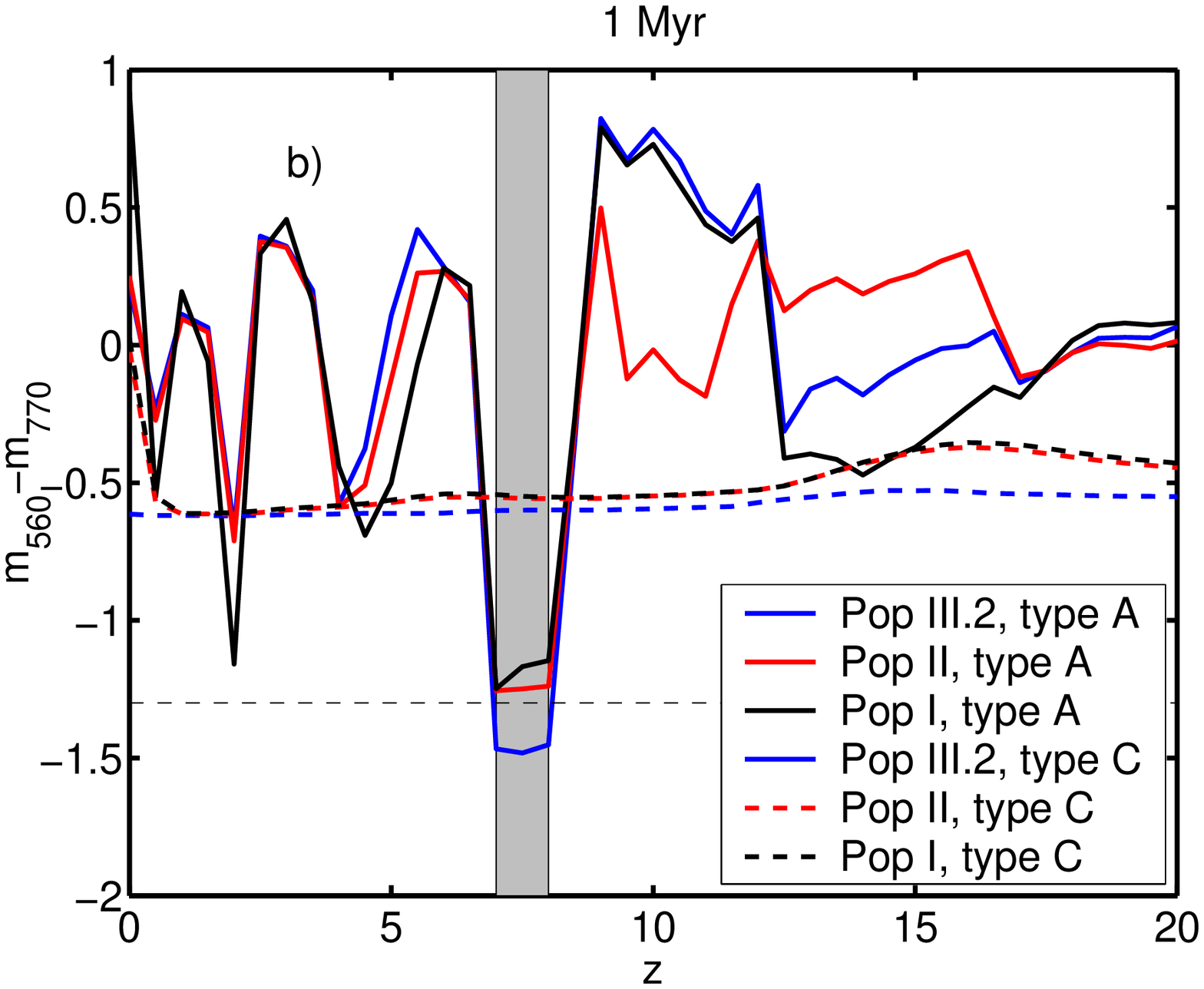}
\caption{The redshift evolution of the {\bf a)} $m_{277}-m_{444}$ and {\bf b)} $m_{560}-m_{770}$ colours of 1 Myr old galaxies. The solid lines represent predictions for type A galaxies: Pop III.2 (blue solid), Pop II (red solid) and Pop I (black solid). The dashed lines represent the corresponding predictions for type C models. A \citet{Kroupa} IMF is assumed for both pop I and pop II objects. Since all pop III objects display very similar colours at this age, only the pop III.2 predictions are included to avoid cluttering. Galaxies with SEDs dominated by nebular emission (type A; solid lines) are seen to display a far more dramatic redshift evolution than those with SEDs dominated by direct star light (type C; dashed lines). In the left panel, the $m_{277}-m_{444}<0$ criterion (thin horizontal dashed line) suggested by \citet{Inoue} for pop III galaxies at $z\approx 8$ is seen to also pick up on many other types of young type C galaxies over a wide range of redshifts. Here, the shaded region indicates the redshift interval $z\approx 6.7$--8.5, over which the \citet{Inoue} criterion holds for our pop III model galaxies. The right panel demonstrates that Pop III, type A galaxies at $z\approx 7$--8 (shaded region) should appear at $m_{560}-m_{770}\leq -1.3$ (thin horizontal dashed line), which is bluer than all other galaxy model in this diagram, and hence provides a cleaner selection of pop III candidates.\label{typeA_zevol}}
\end{figure*}

One major obstacle for {\it any} colour criteria applied to SEDs dominated by nebular emission, is that they evolve very quickly as a function of redshift. In Fig.~\ref{typeA_zevol} we plot the redshift dependence of the $m_{277}-m_{444}$ and $m_{560}-m_{770}$ colours for young (1 Myr old) pop III, II and I galaxies of type A. As seen, both colours evolve dramatically as a function of redshift. By contrast, the colours of equally young type C galaxies (i.e. with SEDs dominated by direct starlight; dashed lines) hardly evolve at all. Even though the $m_{277}-m_{444}<0$ criterion picks up on pop III galaxies at $z\approx 6.7$--8.5, it also selects young pop I and II galaxies of both type A and C at $z<5$ and at $z\geq 6.5$. The risk of confusion can be limited, but not completely removed, by adding photometry in other NIRCam filters and applying drop-out criteria. For instance, additional observations in the F090W and F115W filters should effectively limit the redshift range of the sample to $z\approx 6.5$--9.0 since objects in this redshift range are expected to appear as drop-outs in F090W, but not in F115W, due to IGM absorption shortward of Ly$\alpha$. 

A MIRI-based pop III criterion like $m_{560}-m_{770}\leq -1.3$ (shaded region in Fig.~\ref{typeA_MIRIcrit}) is not likely to generate nearly as many interlopers. This colour is useful as a pop III diagnostic over the limited range $z\approx 7$--8 (shaded region in Fig.~\ref{typeA_zevol}) only, but within that range, pop III galaxies produce bluer colours than {\it all} other kinds of $Z\leq 0.020$, $z>1$ galaxies in our model grid. Even though we have here only plotted the colours of very young galaxies (1 Myr old), this holds for all interloper ages (up to the age of the Universe at each redshift considered).  While there is some room for confusion with $z<1$ galaxies -- which due to dust emission in the F560W filter (not taken into account in our models) may potentially produce similar colours in this diagram -- such objects are not likely to display apparent magnitudes in the same range as high-redshift pop III galaxies. Hence, objects that end up in the upper left corner of Fig.~\ref{typeA_MIRIcolcol}, and have apparent magnitudes in the range expected for high-redshift galaxies, are likely to be $z\approx 7$--8 pop III galaxies even in the absence of additional redshift constraints.

\subsection{Sensitivity issues}
While this scheme would seem to give a cleaner selection of pop III galaxy candidates than that proposed by \citet{Inoue b}, it suffers from one obvious drawback: the lower sensitivity of MIRI implies a minimum mass for detection that is an order of magnitude higher than in the case where only NIRCam filters are used (see Fig.~\ref{Mmin_singlez}). Even in the case of 100 h exposures per filter, the mass converted into stars would need to be on the order of $\sim 10^7\ M_\odot$ to allow detection in the F560W and F770W filters. The maximum mass that stellar populations consisting entirely of pop III stars can reach is unknown, but current simulations suggest that unenriched halos are unlikely to attain total masses in excess of $M\sim 10^8\ M_\odot$ at $z>7$ \citep{Trenti et al.}. To produce $\sim 10^7\ M_\odot$ worth of pop III stars, essentially all of the baryons in a $M\sim 10^8\ M_\odot$ object would need to be converted into stars (somehow evading negative feedback effects) in a limited amount of time (up to $\sim 10^7$ yr), which seems highly unrealistic. By hunting for pop III galaxies behind lensing clusters with magnification $\mu\approx 100$ \citep[MACS J0717.5+3745; e.g.][]{Zitrin et al.,Zackrisson et al. c}, the required stellar population mass can in principle be lowered to $\sim 10^5\ M_\odot$ (implying a star formation efficiency of $\epsilon \sim 10^{-2}$). However, since gravitational lensing reduces the volume probed within a certain redshift interval by a factor equal to the magnification itself, the number density of pop III galaxies would need to be very high to make this strategy successful. Detailed lensing calculations would be required to assess the prospects of endeavours of this type (Zackrisson et al., in prep.). The alternative would be to hunt for pop III galaxies using a coarse set colour criteria based on NIRCam-only data \citep{Inoue b}, accept the possibility of pop I/II interlopers and then do spectroscopic follow-up with JWST/NIRSpec for {\it all} candidates. The exposure times required to detect a $z\approx 8$ pop III object of a given mass in F560W or F770W with MIRI is actually comparable to that required to reach the continuum level at $\approx 4.5\ \mu$m and spectroscopically confirm the absence of a [OIII]$\lambda$5007 line with NIRSpec. Hence, unless the number of candidate pop III galaxies turn out to be very large ($\approx 100$; limited by the number of NIRSpec microshutters) within a given field ($3\arcmin \times 3\arcmin$) at the relevant magnitudes, the NIRCam + NIRSpec strategy may be more economical in terms of telescope time.

\section{The spectral signatures of type C, pop III galaxies}
\label{typeC}
\begin{figure}
\plotone{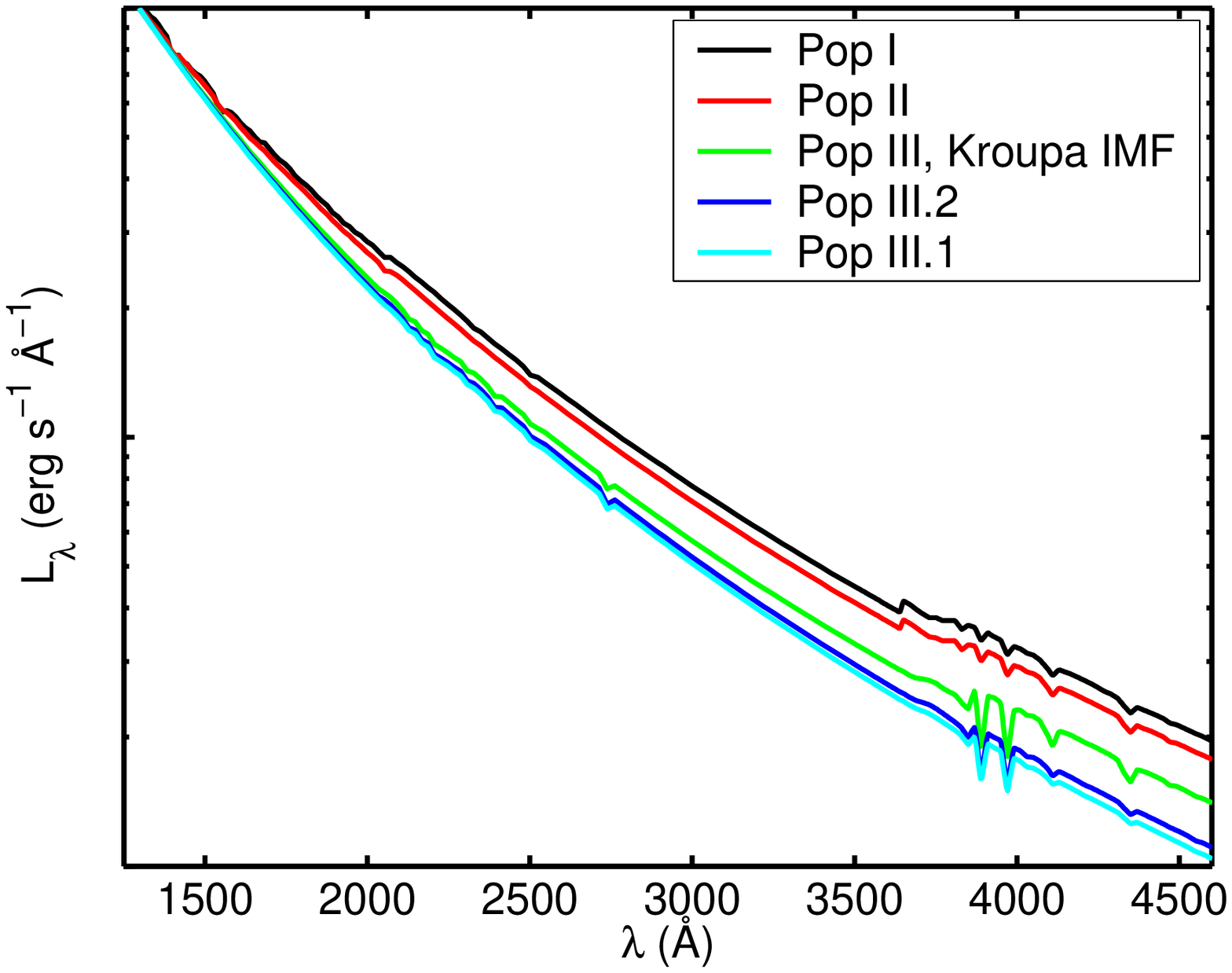}
\caption{The rest frame UV spectra of 1 Myr old, instantaneous-burst pop I (black line), pop II (red), pop III with a \citet{Kroupa} IMF (green), pop III.2 (blue) and pop III.1 (cyan) galaxies of type C. In all cases, we assume that the SED is entirely due to direct star light, i.e. that the leakage of Lyman continuum photons is so overwhelming that the contribution from nebular emission is negligible. The steepness of the spectra is seen to increase when going from high to low metallicities (from black to red and green lines), and when going from a \citet{Kroupa} IMF to top-heavy ones (from green to blue and cyan lines). This suggests that pop III galaxies should be identifiable using broadband filters that measure the restframe UV slope. All spectra have been normalized to the same flux at around 1300 \AA{} to allow differences in the spectral slope to be more easily spotted.  
\label{UVslope}}
\end{figure}

If pop III galaxies can experience sufficient Lyman continuum leakage to render the nebular contributions to their SEDs negligible, such type C objects would stand out in JWST surveys because of their very blue rest-frame UV continuum slopes. This is demonstrated in Fig.~\ref{UVslope}, where we compare the SEDs of 1 Myr old, instantaneous-burst pop III, II and I galaxies at rest-frame wavelengths from 1250 Å to 4600 Å (the range accessible for the NIRCam instrument for a $z=10$ target). The stellar continuum is seen to become progressively steeper when going towards lower metallicities and more top-heavy IMFs -- a fact previously noted by e.g. \citet{Bromm et al. b}, \citet{Schaerer b}, \citet{Tumlinson et al. b}, \citet{Schaerer & Pello}, \citet{Bouwens et al.}, \citet{Taniguchi et al.} and \citet{Raiter et al. b}. However, the difference between the pop III.2 and pop III.1 galaxies appears to be undetectably small. As seen in Fig.~\ref{spectra}, inclusion of nebular emission would make the observed continua substantially redder, thus potentially allowing type C galaxies to be identified by their very blue broadband colours. The filters would, however, need to be carefully selected for the redshift range of interest to prevent small amounts of lingering nebular emission from interfering with this spectral signature.

\begin{deluxetable}{lllll}
\tabletypesize{\scriptsize}
\tablecaption{UV slope$^\mathrm{1}$ $\beta$ for instantaneous-burst models}
\tablewidth{0pt}
\tablehead{
\colhead{Metallicity/IMF} & \colhead{Type$^\mathrm{2}$} & \colhead{1 Myr} & \colhead{10 Myr} & \colhead{100 Myr}
}
\startdata
Pop III.1 & A  & $-2.3$ &  $-$ & $-$ \\
Pop III.1 & C  & $-3.5$ &  $-$ & $-$ \\
Pop III.2 & A  & $-2.2$ &  $-2.7$ & $-2.8$ \\
Pop III.2 & C  & $-3.5$ &  $-3.2$ & $-2.8$ \\
Pop III, \citet{Kroupa}& A & $-2.3$  & $-2.7$ & $-2.6$\\
Pop III, \citet{Kroupa}& C & $-3.4$  & $-3.1$ & $-2.6$\\
Pop II & A & $-2.2$ & $-2.6$  & $-2.0$\\
Pop II & C & $-3.1$ & $-2.7$  & $-2.0$\\
Pop I  & A & $-2.8$ & $-2.1$  & $-1.3$\\
Pop I  & C & $-2.9$ & $-2.1$  & $-1.3$\\
\enddata
\tablenotetext{1}{Derived using $\beta=4.29(J_{125}-H_{160})-2$ at $z=7$ \citep{Bouwens et al.}}
\tablenotetext{2}{Type A: $f_\mathrm{esc}=0$ (maximal nebular emission)\\ Type B: $f_\mathrm{esc}=1$ (purely stellar SED).}
\label{UVslope_table}
\end{deluxetable}

\begin{figure*}
\plottwo{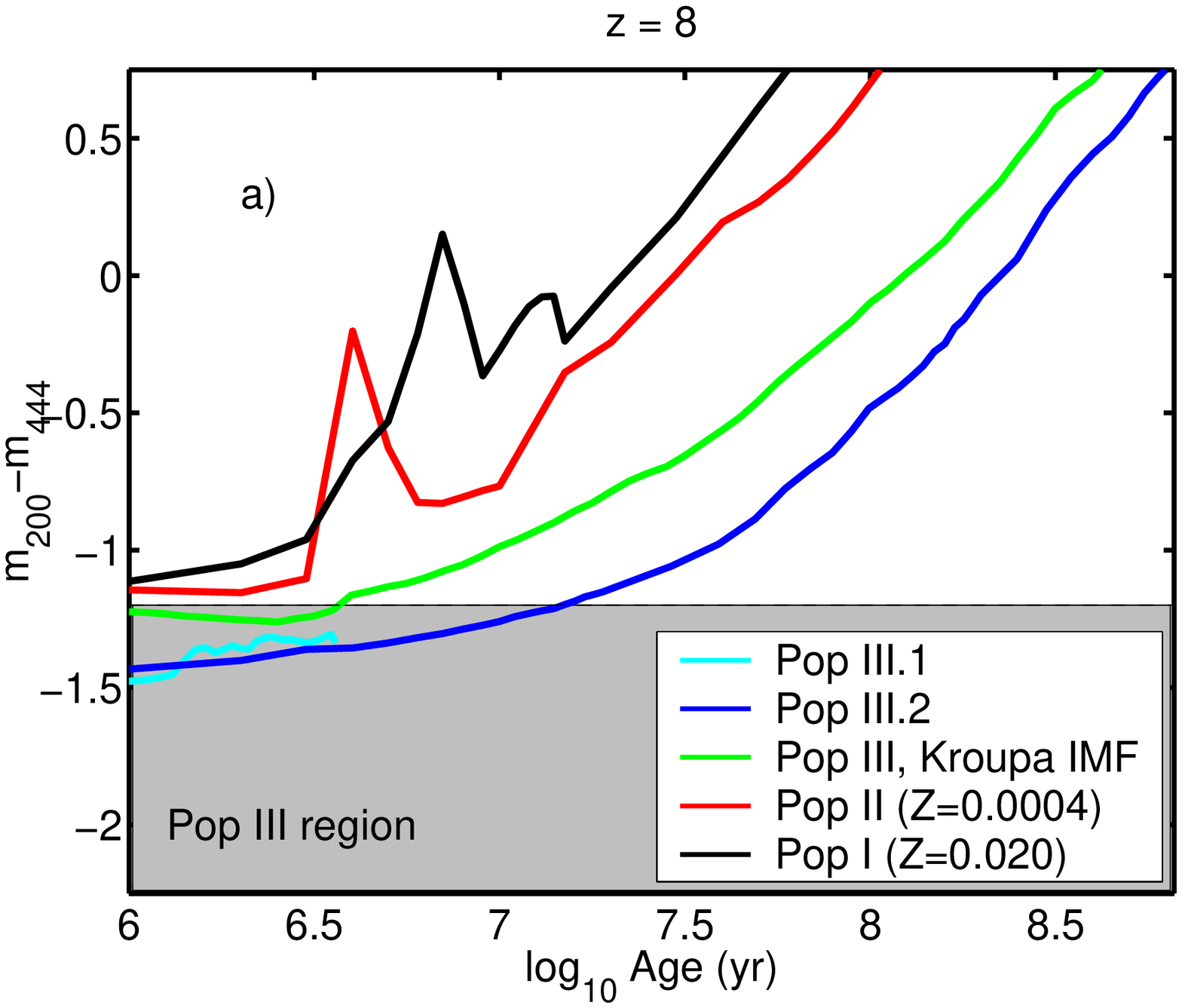}{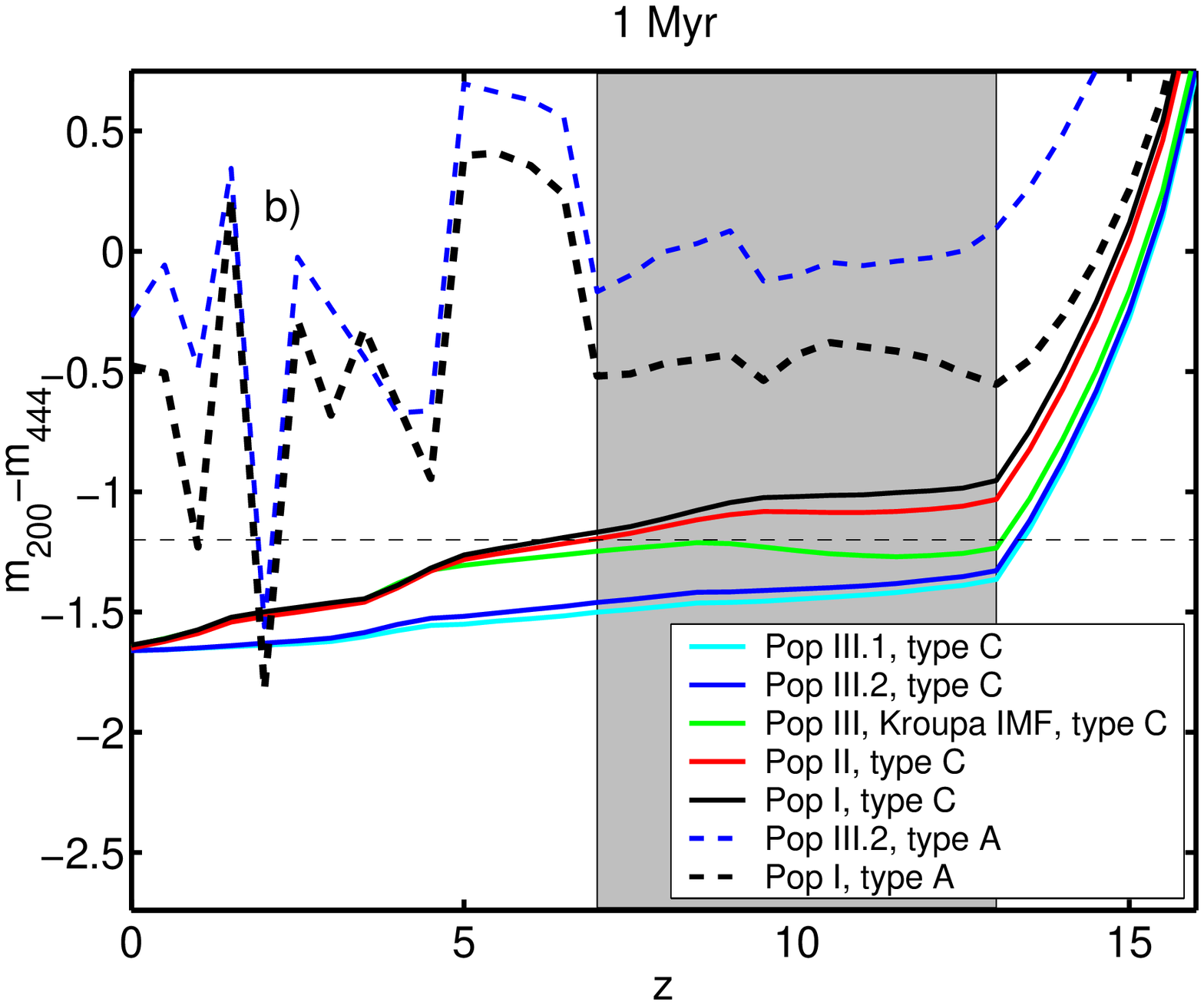}
\caption{The prospects of using the JWST/NIRCam $m_{200}-m_{444}$ colour to distinguish type C, pop III galaxies from other objects. {\bf a)} The $m_{200}-m_{444}$ evolution of instantaneous-burst, type C populations as a function of age at $z=8$. The line colours indicate different metallicity and IMF combinations: Pop III.1 (cyan line), Pop III.2 (blue), Pop III with a \citet{Kroupa} IMF (green), Pop II (red), Pop I (black). A \citet{Kroupa} IMF has been assumed for all of the $Z=0.0004$--0.020 populations. The shaded region indicates the $m_{200}-m_{444} = -1.2$ limit below which only pop III galaxies (green, blue and cyan lines) are likely to be found. All type A galaxies appear at $m_{200}-m_{444} \geq -0.4$ at this redshift. The error bars on the observed colours would be $\approx 0.15$ mag in the case of $10\sigma$ detections in the relevant filters. {\bf b)} The redshift evolution of the $m_{200}-m_{444}$ colours for 1 Myr old populations. The solid lines are the same as in the left panel, whereas the dashed lines represent a selection of type A galaxies: pop III.2 (blue dashed) and pop I (black dashed).
Throughout the redshift range $z\approx 7$--13 (shaded region), the criterion $m_{200}-m_{444}\leq -1.2$ (thin horizontal dashed line) should effectively separate pop III objects from all other types of galaxies.
\label{typeC_colcrit}}
\end{figure*}

Basically, a photometric measurement of the rest-frame UV continuum should ideally use filters with as large wavelength separation as possible. However, since IGM absorption shortward of Ly$\alpha$ likely precludes the use of certain NIRCam filters at $z>6$, we have here selected $m_{200}-m_{444}$, as a trade-off between redshift range and UV slope sensitivity. In Fig.~\ref{typeC_colcrit}, we demonstrate that $m_{200}-m_{444}$ should be able do distinguish type C, pop III galaxies from {\it all} kinds of pop I and II galaxies throughout the redshift range $z\approx 7.5$--13. 

In Fig.~\ref{typeC_colcrit}a, we demonstrate that any type C galaxies at $z=8$ (solid lines) that display $m_{200}-m_{444}\leq -1.2$ is bound to be a pop III, type C system. Nebular emission in type A galaxies would only drive the colour redward. For instance, all newborn ($\approx 1$ Myr old) type A galaxies start off with $m_{200}-m_{444}\geq -0.4$ mag at this redshift. The $\sigma(m_{200}-m_{444})\approx 0.15$ mag photometric error resulting from demanding a $10\sigma$ detection (the detection threshold assumed in Fig.~\ref{Mmin}) in both filters should be sufficent to separate pop III.1 and pop III.2 galaxies from pop II and I, but would make it difficult to distinguish pop III galaxies with a \citet{Kroupa} IMF from pop I or pop II galaxies. The $m_{277}-m_{444}$ colour suggested by \citet{Inoue b} as a diagnostic is not as suitable, since the predicted difference between pop III and pop II/I galaxies is smaller.

In the instantanous-burst models used in Fig.~\ref{typeC_colcrit}a, pop III.1 systems only last for about $\approx 3$ Myr before fading away and Pop III.2 systems will eventually (at ages $\gtrsim$ $10^7$ yr) attain colours redder than the $m_{200}-m_{444}\leq -1.2$ criterion. Pop III galaxies with a \citet{Kroupa} IMF evolve redward of this limit after about 4 Myr. More extended star formation scenarios would in principle allow pop III galaxies to retain their $m_{200}-m_{444}\leq -1.2$ colours for longer periods of time. In the case of a constant star formation rate, type C Pop III galaxies with a \citet{Kroupa} IMF would display $m_{200}-m_{444}\leq -1.2$ for about 6 Myr, and Pop III.2 and pop III.1 would meet the same colour criteria for as long as star formation remains active. Of course, a type C object requires that much of the gas of the system has been vacated, and this is likely to prevent pop III star formation from progressing efficiently. Hence, it seems unlikely that type A galaxies can retain their unique UV-slope signatures for much longer than $\sim 10$ Myr.

For reference, we have in Table.~\ref{UVslope_table} included the UV slope $\beta$ ($f_\lambda \propto \lambda^\beta$) predicted for our different pop III, II and I models, derived the way this is currently done for galaxies at $z\approx 7$ \citep[e.g.][]{Bouwens et al.,Dunlop et al.}. Here, $\beta$ is based on the {\it Hubble Space Telescope} WFC3 $J_{125}-H_{160}$ colour, using the \citet{Bouwens et al.} calibration. At $z=7$, these filters sample the rest frame UV at wavelengths of $\approx  1370$--1750 \AA{} and $\approx 1760$--2110 \AA{}, respectively. We caution, however, that the $\beta$ parameter does depend on the exact wavelength range sampled \citep[e.g.][]{Raiter et al. b}, especially for models involving nebular emission. Additional UV slope predictions for some of the pop III SEDs used in this work are available in electronic format, along with numerous other quantities, from \citet{Raiter et al. b}. 
 
\subsection{The redshift evolution of type C galaxy colours}
Since the type C SEDs displayed in Fig.~\ref{UVslope} show some deviations from pure power-law spectra, colours such as $m_{200}-m_{444}$ will evolve as a function of redshift. In Fig.~\ref{typeC_colcrit}b, we show the redshift dependence of $m_{200}-m_{444}$ for 1 Myr old galaxies of both type A and C. Throughout the redshift range $z\approx 7$--13 (shaded region), the region $m_{200}-m_{444}\leq -1.2$ should effectively separate pop III objects from all other types of galaxies. All type A galaxies (dashed lines) appear at $m_{200}-m_{444}\geq -0.6$ within this redshift range. The redshift limits over which the $m_{200}-m_{444}\leq -1.2$ criterion applies are set by the fact that, at $z<7$, pop I and pop II of type C start to occupy this region of colour space, and at $z>13$, Ly$\alpha$ absorption in the intergalactic medium is expected to cause substantial flux losses in the F200W filter, thereby rendering the $m_{200}-m_{444}$ colour extremely red for all populations. Dust extinction, if effective in pop III galaxies, would of course make the $m_{200}-m_{444}$ colour redder and could shroud some bona fide pop III systems in the colours of more metal-enriched systems. While it is possible that some pop III galaxies could masquerade as pop II/I galaxies (either due to old age or due to dust), the opposite is not true -- if objects are detected at $m_{200}-m_{444}\leq -1.2$, they would be very good pop III candidates.  
 
\subsection{Lingering nebular emission} 
While measurements of the UV slope of type C galaxies could in principle help constraint the pop III IMF, the question remains whether pure type C, pop III galaxies are at all likely to be detectable. As explained in Sect.~\ref{scenarios}, it would take a tremendous amount of leakage to bring nebular emission down to negligible levels in the SED of a pop III galaxy. Due to the very red colours of young type A galaxies ($m_{200}-m_{444}\geq -0.4$) and the high ratio of nebular to stellar flux in the F444W filter (Fig.~\ref{fneb}), even small amounts of lingering nebular light would jeopardize the possibility of identifying pop III, type C objects. We find that, depending on the exact redshift ($z\approx 7$--13), IMF and age, a Lyman continuum escape fraction of $f_\mathrm{esc}\gtrsim 0.95$--0.99 would be required for a pop III galaxy to meet the $m_{200}-m_{444}\leq -1.2$ limit. With such small nebular contributions to the overall flux, a type C pop II galaxy would need to be an order of magnitude more massive than a type A object to be detectable with JWST (Fig.~\ref{Mmin}). While the relative contribution from nebular emission decreases at high ages for instantaneous burst models ($\gtrsim 10$ Myr for IMFs that allow populations to remain luminous for that long), the intrinsic UV slope at the same time becomes redder (see Fig.~\ref{typeC_colcrit}a), thereby jeopardizing any chances of detecting a unique pop III signature. Adopting a more extended star formation history would make pop III galaxies retain their very blue UV slopes for longer, but also keeps the $f_\mathrm{neb}/f_\mathrm{stars}$ ratio up, and therefore still requires a very high $f_\mathrm{esc}$ to allow the broadband criteria to apply. Switching to colours based on JWST filters at shorter central wavelengths would reduce the relative contribution from nebular emission somewhat (this is evident from Fig.~\ref{spectra}), but also decreases the difference in magnitudes between the intrinsic colours of pop III and pop II/I. For instance, pop III galaxies at $z=8$ could display $m_{150}-m_{200}$ colours (based on fluxes in the NIRCam F150W and F200W filters at 1.5 and 2.0 $\mu$m, repspectively) redder than those of pop I and II galaxies for $\sim 10^7$ yr even in the case of $f_\mathrm{esc}\approx 0.9$ (i.e. lower than the $f_\mathrm{esc}\gtrsim 0.95$--0.99 quoted above, albeit not by much), but the colour difference between these models is then $\leq 0.05$ mag, which would be very difficult to measure in practice.

Whether pop III objects with Lyman-continuum escape fractions $f_\mathrm{esc}\approx 0.9$--0.99 exist is an open question. Recent simulations by \citet{Johnson et al. b} suggest that $f_\mathrm{esc}$ in this range may be produced if the pop III IMF is extremely top-heavy, but taken at face value, the prospects of identifying pop III galaxies through the colour signatures of type A objects appear more promising, since they are likely to hold for a wider range of escape fractions ($f_\mathrm{esc}\approx 0$--0.5). 

\section{Discussion}
\label{discussion}

\subsection{Caveats related to the [OIII] method}
In Sect.~\ref{typeA}, we argued that pop III galaxies with SEDs dominated by nebular emission (type A) could potentially be identified based on their JWST colours, since a combination of strong hydrogen emission lines like H$\alpha$, yet absent metal emission lines like [OIII]$\lambda$5007 would gives rise to very peculiar colours over certain redshift intervals. Here, we discuss a couple of potential caveats with this method.

\subsubsection{The oxygen abundances of pop III galaxies}
Our tests indicate that pop III, type A signatures based on the lack of [OIII]$\lambda$5007 emission are erased as soon as the oxygen-to-hydrogen abundance relative to that of the Sun reaches [O/H] $\geq -2.2$ (or (O/H) $\geq 5\times 10^{-6}$ in absolute numbers). Depending on the metallicity threshold at which gas switches from the formation of pop III stars with a top-heavy IMF to pop II with an IMF more typical of that in the local Universe \citep[often assumed to happen at $Z=10^{-4}$ to $10^{-6}$, e.g.][]{Schneider et al. b}, and on the oxygen yields of the first supernovae, this oxygen criterion may translate into a metallicity criterion for type A, pop III galaxy signatures that is either slightly lower or higher than that which governs the pop III-II transition. Hence, the method proposed by us and \citet{Inoue b} for detecting pop III, type A galaxies may in principle either miss a substantial fraction of galaxies that are still able to form pop III stars, or to catch galaxies that have already switched to pop II star formation. Clearly, a method of this type is only meant to generate pop III galaxy {\it candidates}. Follow-up spectroscopy will be required to further probe the exact nature of these targets. For objects that are already close to the detection threshold for imaging in ultra deep fields, this may of course be very difficult to achieve with JWST itself. However, such follow-up observations may for instance be feasible with the upcoming 42 m European Extremely Large Telescope\footnote{http://www.eso.org/sci/facilities/eelt/}.

\subsubsection{Ionization levels}
Another caveat associated with weak or absent [OIII]$\lambda$5007 emission as a criterion for pop III galaxies is that pop II/I galaxies with very low ionization parameters may also display very low [OIII] fluxes. In the framework of a single spherical nebulae surrounding the star-forming core of the galaxy, this could correspond to the situation where stellar feedback has increased the inner radius of the cloud to large radii. However, such objects may no longer qualify as type A objects, but may rather belong to type B (see Fig.~\ref{schematic}), with a drastically lower nebular fluxes and potentially distinct spectral characteristics as a result. A detailed investigation of the properties of type B galaxies would be required to settle this issue.

\subsection{Narrowband imaging}
One may, perhaps, argue that narrowband surveys are more likely to be successful in isolating pop III galaxies than any broadband survey. Aside from ``classical'' narrowband features of pop III objects, like very high Ly$\alpha$ or [HeII]$\lambda1640$ equivalent widths, one may for instance look for the Balmer jump at $0.36\ \mu$m \citep[e.g.][]{Schaerer a,Schaerer b,Inoue b}. All objects with spectra dominated by nebular emission should display a sudden jump in the continuum level at the short-wavelength side compared to the long wavelength side of this limit. Since metal-enriched stars are less efficient in ionizing hydrogen and have redder stellar continua, they will -- relatively speaking -- be less dominated by nebular emission (see Fig.~\ref{fneb}). Therefore, one may conjecture that the objects with the most pronounced Balmer discontinuities are likely to be pop III galaxies, and that suitably placed narrowband filters would be able to pick up on this. However, our models indicate that the difference between the Balmer jumps between pop II and pop III galaxies are so small that it will be next to impossible to tell them apart using this feature. While the Balmer jump may still be used for identifying extremely young, nebular-dominated objects at low metallicities (albeit not necessarily just pop III objects), the redshift range over which this feature can be efficiently probed with a given setup of JWST narrowband filters is very small ($\Delta z \sim 0.1$). This makes the number of potential targets in the corresponding volume so tiny, that it would seem more time-efficient to use broadband criteria to pre-select pop III galaxy candidates and then to go directly for follow-up spectroscopy.

\subsection{How much do the details of the pop III IMF matter?}
\citet{Bromm et al. b} argue that very massive pop III stars are self-similar, in the sense that a fixed mass of gas converted into either $100\ M_\odot$, $300\ M_\odot$ or $500\ M_\odot$ stars would produce nearly identical SEDs. Indeed, the lifetimes, ZAMS temperatures and the time-averaged, mass-normalized Lyman continuum fluxes of $M\approx 100$-500 $M_\odot$ stars differ by no more than 10--30 \% \citep{Schaerer a}. However, the approximation of self-similarity deteriorates once pop III stars at $M<100\ M_\odot$ are considered, and does not hold for the HeII ionizing spectrum and the corresponding recombination lines, or for stars that have evolved off the main sequence, as discussed in detail by \citet{Schaerer a}.

As seen in Fig.~\ref{Mmin}, the mass-to-light ratios produced by our pop III.1 ($\sim 100\ M_\odot$) and pop III.2 ($\sim 10\ M_\odot$) IMFs differ by a factor of $\approx 2$--3. This is mostly due to the inclusion of low-mass stars which contribute substantially to the mass and not to the luminosity. Hence, the colours produced by the different pop III IMFs remain very similar (Figs.~\ref{typeA_Inoue_crit}, \ref{typeA_MIRIcrit}, \ref{typeA_MIRIcolcol} and \ref{typeC_colcrit}). For a fixed star formation history, galaxies with pop III.2 IMFs can continue to shine for longer than those with pop III.1 IMFs, but their colours are otherwise identical. Pop III galaxies with \citet{Kroupa} IMFs display slightly less extreme colours, but this is largely caused by a shift in the time scales -- pop III.2 galaxies will display the same colours, just at a slightly higher age. It is only in the case of a purely stellar SED (our case C; Fig.~\ref{typeC_colcrit}) that Pop III.1 and pop III.2 get separated  from the  \citet{Kroupa} IMF even at very low ages. Hence, the detection of objects with extreme colours would support the idea of a pop III IMF with more high-mass stars than predicted by the \citet{Kroupa} IMF, but the prospects of disentangling a pop III.1 IMF from a pop III.2 IMF using integrated colours alone seem bleak.  

\section{Summary}
\label{summary}
Yggdrasil is a new spectral synthesis model which includes pop I, II and III stars, dark stars, nebular emission and dust extinction. Using this model, we derive the stellar population masses of the faintest galaxies detectable by JWST through broadband imaging in ultra deep fields. Assuming 100 h exposures per filter, we find that JWST may be able to detect pop III galaxies with stellar masses as low as $\sim 10^5\ M_\odot$ at $z=10$, whereas the corresponding limit for pop II/I galaxies is $\sim 10^6 M_\odot$. We also argue that the broadband fluxes of young (age $\lesssim 10$ Myr) pop III galaxies are likely to be significantly affected by nebular emission, unless the fraction of ionizing radiation escaping directly into the intergalactic medium is extremely high ($f_\mathrm{esc}\gtrsim 0.95$). 

We also discuss different strategies for selecting Pop III galaxy candidates based on their JWST broadband colours, both in the limiting cases of having dominant (low $f_\mathrm{esc}$) and negligible (very high $f_\mathrm{esc}$) nebular emission. In the former case, we particularly highlight the possibility of adding imaging in two MIRI filters (F560W and F770W) to the NIRCam filter sets usually discussed in the context of JWST ultra deep fields, since this should allow for a clean selection of pop III galaxies at $z\approx 7$--8. Selecting targets for spectroscopic follow-up using colour criteria based solely on NIRCam filters would be more economical in terms of exposure time, but also increases the risk of interlopers. In the case of pop III galaxies dominated by direct star light (very high $f_\mathrm{esc}$), we argue that imaging in the NIRCam F200W and F444W filters should allow pop III candidates to be selected throughout the redshift range $z\approx 6.5$--13.

\acknowledgments
E. Zackrisson, C.-E. Rydberg and G. \"Ostlin acknowledge a grant from the Swedish National Space Board and the Swedish Research Council. D. Schaerer is supported by the Swiss National Science Foundation. The authors are indepted to Marcia J. Rieke and Kay Justtanont for giving us access to the NIRCam and MIRI broadband filter profiles prior to public release. The anonymous referee is thanked for useful comments which helped to improve the quality of the paper.\vspace{5mm}


\begin{thebibliography}{}
\bibitem[Anders \& Fritze-v. Alvensleben(2003)]{Anders & Fritze-v. Alvensleben}
Anders, P.; Fritze-v. Alvensleben, U. 2003, A\&A, 401, 1063 
\bibitem[Beers \& Christlieb(2005)]{Beers & Christlieb}
Beers, T. C., Christlieb, N 2005, ARA\&A, 43, 531 
\bibitem[Bouwens et al.(2010)]{Bouwens et al.}
Bouwens, R. J., et al. 2010, ApJ, 708, L69
\bibitem[Bromm et al.(1999)]{Bromm et al. a}
Bromm, V., Coppi, P. S., Larson, R. B. 1999, ApJ, 527, L5 
\bibitem[Bromm et al.(2001)]{Bromm et al. b}
Bromm, V., Kudritzki, R. P., Loeb, A. 2001, ApJ, 552, 464
\bibitem[Bromm \& Larson(2004)]{Bromm & Larson}
Bromm, V., \& Larson, R. B. 2004, ARA\&A, 42, 79
\bibitem[Cai et al.(2011)]{Cai et al.}
Cai, Z., et al. 2011, ApJ, 736 L28
\bibitem[Calzetti(1997)]{Calzetti} 
Calzetti, D. 1997b, in AIP Conf. Proc. 408, The Ultraviolet Universe at Low
and High Redshift, eds. W. H. Waller, M. N. Fanelli, J. E. Hollis, \& A. C. Danks (Woodbury:AIP), 403 (arXiv:astro-ph/9706121)
\bibitem[Clark et al.(2011)]{Clark et al.}
Clark, P. C., Glover, S. C. O., Klessen, R. S., Bromm, V. 2011, ApJ, 727, 110
\bibitem[Dawson et al.(2004)]{Dawson et al.}
Dawson, S., et al. 2004, ApJ, 617, 707
\bibitem[Dijkstra \& Wyithe(2007)]{Dijkstra & Wyithe}
Dijkstra, M., \& Wyithe, J., S. B., 2007, MNRAS, 379, 1589
\bibitem[Dunlop et al.(2011)]{Dunlop et al.}
Dunlop, J. S., McLure, R. J., Robertson, B. E., Ellis, R. S., Stark, D. P., Cirasuolo, M., de Ravel, L. 2011, MNRAS, submitted (arXiv1102.5005)
\bibitem[Ferland et al.(1998)]{Ferland et al.}
Ferland, G. J., Korista, K. T., Verner, D.A., Ferguson, J.W., Kingdon, J.B., Verner, E.M. 1998, PASP, 110, 761
\bibitem[di Serego Alighieri et al.(2008)]{di Serego Alighieri} 
di Serego Alighieri, S., Kurk, J., Ciardi, B., Cimatti, A., Daddi, E., Ferrara, A. 2008, In: Low-Metallicity Star Formation: From the First Stars to Dwarf Galaxies, Proceedings of the International Astronomical Union, IAU Symposium, Volume 255, p. 75
\bibitem[Fontana et al.(2010)]{Fontana et al.}
Fontana, A., et al. 2010, ApJ, 725 L205
\bibitem[Fosbury et al.(2003)]{Fosbury et al.}
Fosbury, R. A. E., et al. 2003, ApJ, 596, 797
\bibitem[Gardner et al. (2006)]{Gardner et al.}
Gardner, J. P., et al. 2006, Space Sci.Rev. 123, 485
\bibitem[Gnedin et al. (2008)]{Gnedin et al.}
Gnedin, N. Y., Kravtsov, A. V., Chen, H.-W. 2008, ApJ, 672, 765
\bibitem[Greif \& Bromm(2006)]{Greif & Bromm}
Greif, T. H., \& Bromm, V. 2006, MNRAS, 373, 128
\bibitem[Greif et al.(2007)]{Greif et al. a}
Greif T. H., Johnson J. L., Bromm V., Klessen R. S., 2007, ApJ,
670, 1
\bibitem[Greif et al.(2009)]{Greif et al. b}
Greif, T. H., Johnson, J. L., Klessen, R. S., Bromm, V. 2009, MNRAS, 399, 639
\bibitem[Greif et al.(2010)]{Greif et al. c}
Greif, T. H., Glover, S. C. O., Bromm, V. Klessen, R. S. 2010, ApJ, 716, 510
\bibitem[Greif et al.(2011a)]{Greif et al. d}
Greif, T. H., Springel, V., White, S. D. M., Glover, S. C. O., Clark, P. C., Smith, R. J., Klessen, R. S., Bromm, V. 2011, ApJ, submitted (arXiv1101.5491)
\bibitem[Greif et al.(2011b)]{Greif et al. e}
Greif, T. H., White, S. D. M., Klessen, R. S., Springel, V. 2011, ApJ, submitted (arXiv1101.5493)
\bibitem[Hayes et al.(2011)]{Hayes et al.} 
Hayes, M., Schaerer, D., \"Ostlin, G., Mas-Hesse, J. M., Atek, H., Kunth, D. 2011, ApJ, 730, 8 
 \bibitem[Heger et al.(2002)]{Heger et al.}
Heger, A., Woosley, S., Baraffe, I. \& Abel, T. 2002, in Lighthouses of the Universe: The Most Luminous Celestial Objects and Their Use for Cosmology,  eds. M. Gilfanov, R. Sunyaev, and E. Churazov, ESO Astrophysics symposia, Springer-Verlag, p. 369 (astro-ph/0112059)
\bibitem[Inoue(2010)]{Inoue}
Inoue, A. K. 2010, MNRAS, 401, 1325 
\bibitem[Inoue et al.(2011a)]{Inoue et al.}
Inoue, A. K., et al. 2011, MNRAS, 411, 2336
\bibitem[Inoue(2011b)]{Inoue b}
Inoue, A. K. 2011, MNRAS, in press (arXiv1102.5150)
\bibitem[Jimenez \& Haiman(2006)]{Jimenez & Haiman}
Jimenez R. \& Haiman Z., 2006, Nature, 440, 501 
\bibitem[Johnson(2010)]{Johnson} 
Johnson, J. L. 2010, MNRAS, 404, 1425
\bibitem[Johnson, Greif \& Bromm(2008)]{Johnson et al. a} 
Johnson, J. L., Greif, T. H., Bromm, V., 2008, MNRAS, 388, 26 
\bibitem[Johnson et al.(2009)]{Johnson et al. b} 
Johnson, J. L., Greif, T. H., Bromm, V., Klessen, R. S., Ippolito, J. 2009, MNRAS, 399, 37
\bibitem[Karlsson et al.(2008)]{Karlsson et al. a}
Karlsson, T., Johnson, J. L., Bromm, V. 2008, ApJ, 679, 6
\bibitem[Karlsson et al.(2011)]{Karlsson et al. b}
Karlsson, T., Bromm, V., Bland-Hawthorn, J. 2011, Rev. Mod. Phys., submitted (arXiv1101.4024)
\bibitem[Kewley \& Dopita(2002)]{Kewley & Dopita}
Kewley, L. J.; Dopita, M. A. 2002, ApJS, 142, 35
\bibitem[Kitayama et al.(2004)]{Kitayama et al.}
Kitayama, T., Yoshida, N., Susa, H., Umemura, M. 2004, ApJ 613, 631
\bibitem[Kroupa(2001)]{Kroupa} 
Kroupa, P. 2001, MNRAS, 322, 231
\bibitem[Leitherer et al.(1999)]{Leitherer et al.} 
Leitherer C., et al. 1999, ApJS, 123, 3
\bibitem[Mackey et al.(2003)]{Mackey et al.} 
Mackey, J., Bromm, V., Hernquist, L. 2003, ApJ, 586, 1
\bibitem[Maio et al.(2011)]{Maio et al.}
Maio, U., Koopmans, L. V. E., Ciardi, B. 2011, MNRAS,  412, L40
\bibitem[Malhotra \& Rhoads(2002)]{Malhotra & Rhoads}
Malhotra, S., Rhoads, J. E. 2002, ApJ, 565, L71
\bibitem[Maraston(2005)]{Maraston} 
Maraston, C. 2005, MNRAS, 362, 799
\bibitem[Nagao et al.(2005)]{Nagao et al. a}
Nagao, T., et al. 2005, ApJ, 631, L5
\bibitem[Nagao et al.(2006)]{Nagao et al. b}
Nagao, T., Maiolino, R., \& Marconi, A. 2006, A\&A, 459, 85
\bibitem[Nagao et al.(2008)]{Nagao et al. c}
Nagao, T., et al. 2008, ApJ, 680, 100
\bibitem[Nakamura \& Umemura(2002)]{Nakamura & Umemura} 
Nakamura, F., \& Umemura, M. 2002, ApJ, 569, 549 
\bibitem[Oh et al.(2001)]{Oh et al.}
Oh, S. P., Haiman, Z., \& Rees, M. J. 2001, ApJ, 553, 73 
\bibitem[Ohkubo et al.(2009)]{Ohkubo et al.}
Ohkubo, T., Nomoto, K., Umeda, H., Yoshida, N., Tsuruta, S. 2009, ApJ, 706, 1184 
\bibitem[Ouchi et al.(2008)]{Ouchi et al.}
Ouchi, M., et al. 2008, ApJS, 176, 301
\bibitem[Panagia(2003)]{Panagia}
Panagia, N. 2003, ChJAS, 3, 115
\bibitem[Pawlik et al.(2011)]{Pawlik et al.}
Pawlik, A. H., Milosavljevi\'c, M., Bromm, V. 2011, ApJ, 731, 54
\bibitem[Pei(1992)]{Pei}
Pei, Y. C. 1992, ApJ, 395, 130
\bibitem[Raiter et al.(2010a)]{Raiter et al. a}
Raiter, A., Fosbury, R. A. E. Teimoorinia, H. 2010, A\&A, 510, 109
\bibitem[Raiter et al.(2010b)]{Raiter et al. b}
Raiter, A., Schaerer, D., Fosbury, R. A. E. 2010 A\&A 523, 64
\bibitem[Razoumov \& Sommer-Larsen(2010)]{Razoumov & Sommer-Larsen}
Razoumov, A. O., \& Sommer-Larsen, J. 2010, ApJ, 710, 1239
\bibitem[Rydberg, Zackrisson \& Scott(2010)]{Rydberg et al.}
Rydberg, C.-E., Zackrisson, E., Scott, P. 2010, in Cosmic Radiation Fields: Sources in the early Universe, ed. M. Raue, T. Kneiske, D. Horns, D. Elsaesser, \& P. Hauschildt, p.26 (arXiv1103.1377)
\bibitem[Salvadori et al.(2007)]{Salvadori et al.}
Salvadori, S., Schneider, R., Ferrara, A. 2007, MNRAS, 381, 647
\bibitem[Salvaterra, Ferrara \& Dayal(2010)]{Salvaterra et al.}
Salvaterra, R., Ferrara, A., \& Dayal, P. 2010, MNRAS, in press (arXiv1003.3873)
\bibitem[Scannapieco et al.(2003)]{Scannapieco et al.}
Scannapieco, E., Schneider, R., \& Ferrara, A. 2003, ApJ, 589, 35
\bibitem[Schaerer(2002)]{Schaerer a}
Schaerer, D. 2002, A\&A, 382, 28
\bibitem[Schaerer(2003)]{Schaerer b}
Schaerer, D. 2003, A\&A, 397, 527
\bibitem[Schaerer \& de Barros(2009)]{Schaerer & de Barros a}
Schaerer, D., \& de Barros, S. 2009, A\&A, 502, 423
\bibitem[Schaerer \& de Barros(2010)]{Schaerer & de Barros b}
Schaerer, D., \& de Barros, S. 2010, A\&A, 515, 73
\bibitem[Schaerer \& Pell\'o(2005)]{Schaerer & Pello} 
Schaerer, D. Pell\'o, R. 2005, MNRAS, 362, 1054 
\bibitem[Schneider et al.(2006a)]{Schneider et al. a}
Schneider, R., Salvaterra, R., Ferrara, A., Ciardi, B. 2006, MNRAS, 369, 825
\bibitem[Schneider et al.(2006b)]{Schneider et al. b}
Schneider, R., Omukai, K., Inoue, A. K., Ferrara, A. 2006, MNRAS, 369, 1437
\bibitem[Sokasian et al.(2004)]{Sokasian et al.}
Sokasian, A., Yoshida, N., Abel, T., Hernquist, L., \& Springel, V. 2004, MNRAS, 350, 47
\bibitem[Spolyar et al.(2008)Spolyar, Freese \& Gondolo]{Spolyar et al.}
Spolyar, D., Freese, K., \& Gondolo, P. 2008, \prl, 100, 051101
\bibitem[Stacy et al.(2008)]{Stacy et al.}
Stacy, A., Greif, T. H., Bromm, V. 2008, MNRAS, 403, 45
\bibitem[Stiavelli \& Trenti(2010)]{Stiavelli & Trenti} 
Stiavelli, M., Trenti, M. 2010, ApJ, 716, L190 
\bibitem[Tan \& McKee(2004)]{Tan & McKee}
Tan, J. C., McKee, C. F. 2004, ApJ, 603, 383
\bibitem[Taniguchi et al.(2010)]{Taniguchi et al.} 
Taniguchi, Y., Shioya, Y., Trump, J. R. 2010, ApJ, 724, 1480
\bibitem[Tegmark et al.(1997)]{Tegmark et al.}
Tegmark, M., Silk, J., Rees, M. J., Blanchard, A., Abel, T., Palla, F. 1997, ApJ, 474, 1
\bibitem[Tornatore, Ferrara \& Schneider(2007)]{Tornatore et al.}
Tornatore, L., Ferrara, A., Schneider, R. 2007, MNRAS, 382, 945
\bibitem[Trenti \& Shull(2010)]{Trenti & Shull}
Trenti, M., Shull, J. M. 2010, ApJ, 712, 435
\bibitem[Trenti et al.(2009)]{Trenti et al.}
Trenti, M., Stiavelli, M., Shull, M. J. 2009, ApJ, 700, 1672 
\bibitem[Trenti \& Stiavelli(2007)]{Trenti & Stiavelli a}
Trenti, M., \& Stiavelli, M. 2007, ApJ, 667, 38
\bibitem[Trenti \& Stiavelli(2009)]{Trenti & Stiavelli b}
Trenti, M., \& Stiavelli, M. 2009, ApJ, 694, 879
\bibitem[Tseliakhovich \& Hirata(2011)]{Tseliakhovich & Hirata}
Tseliakhovich, D., Hirata, C. 2010, PhRvD, 82, 3520
\bibitem[Tumlinson \& Shull(2000)]{Tumlinson & Shull}
Tumlinson, J., \& Shull, J. M. 2000, ApJ, 528, L65
\bibitem[Tumlinson et al.(2001)]{Tumlinson et al. a}
Tumlinson, J., Giroux, M. L., \& Shull, J. M. 2001, ApJ, 550, L1
\bibitem[Tumlinson et al.(2003)]{Tumlinson et al. b}
Tumlinson, J., Shull, J. M., \& Venkatesan, A. 2003, ApJ, 584, 608
\bibitem[Tumlinson(2010)]{Tumlinson}
Tumlinson, J. 2010, ApJ, 708, 1398
\bibitem[V\'azquez \& Leitherer(2005)]{Vazquez & Leitherer}
V\'azquez, G. A., \& Leitherer, C. 2005, ApJ, 621, 695
\bibitem[Weinmann \& Lilly(2005)]{Weinmann & Lilly}
Weinmann, S. M., \& Lilly, S. J. 2005, ApJ, 624, 526
\bibitem[Whalen \& Fryer(2010)]{Whalen & Fryer} 
Whalen, D. J., Fryer, C. 2010, In: Deciphering the Ancient Universe with Gamma-Ray Bursts, AIP Conference Proceedings, Volume 1279, pp. 116-119 (arXiv1009.2543) 
\bibitem[Willott et al.(2007)]{Willott et al.}
Willott, C. J., et al. 2007, AJ, 134, 2435 
\bibitem[Yajima et al.(2011)]{Yajima et al.}
Yajima, H., Choi, J.-H., Nagamine, K. 2011, MNRAS, 412, 411
\bibitem[Yoshida et al.(2003)]{Yoshida et al.}
Yoshida, N., Abel, T., Hernquist, L., \& Sugiyama, N. 2003, ApJ, 592, 645 
\bibitem[Zackrisson et al.(2001)]{Zackrisson et al. a}
Zackrisson, E., Bergvall, N., Olofsson, K., \& Siebert, A. 2001, A\&A, 375, 814
\bibitem[Zackrisson et al.(2008)]{Zackrisson et al. b}
Zackrisson, E., Bergvall, N., Leitet, E. 2008, ApJL, 676, 9
\bibitem[Zackrisson et al.(2010)]{Zackrisson et al. c}
Zackrisson, E., et al. 2010, ApJ, 717, 257
\bibitem[Zitrin et al.(2009)]{Zitrin et al.}
Zitrin, A., Broadhurst, T., Rephaeli, Y., Sadeh, S. 2009, ApJ 707, L102
\end{thebibliography}
\end{document}